\documentclass[aps,prb,showkeys,superscriptaddress,notitlepage]{revtex4-1}
\usepackage[a4paper,top=2.00cm,bottom=2.00cm,left=2.00cm,right=2.00cm]{geometry}
\usepackage{amsfonts}
\usepackage{amsmath}
\usepackage{amssymb}
\usepackage{graphicx}
\usepackage{hyperref}
\usepackage{fleqn}
\newcommand{\be}{\begin{equation}}
\newcommand{\ee}{\end{equation}}
\newcommand{\ber}{\begin{eqnarray}}
\newcommand{\eer}{\end{eqnarray}}
\begin{document}
\title{\vspace*{-1.5cm}
\noindent{\footnotesize\sf Published in the NATO ASI series on ``The Vortex State'', N. Bontemps, Y. Bruynseraede, G. Deutscher, and A. Kapitulnik, eds., Kluwer Academic Press, Dordrecht, The Netherlands, 1994. ISBN: 978-94-010-4422-6.
}\\
\vspace*{1.5cm}
Fermi-Liquid Theory of Non-S-Wave Superconductivity}
\author{Paul Muzikar}
\affiliation{Department of Physics, Purdue University, 
             West Lafayette, IN 47907 USA}
\author{D. Rainer}
\affiliation{Physikalisches Institut, Universit\"at Bayreuth, D-95440 Bayreuth, Germany}
\author{J. A. Sauls}
\affiliation{Department of Physics \& Astronomy, Northwestern University, 
             Evanston, IL 60208 \& Nordita, DK-2100 Copenhagen \O Denmark}
\date{January 13, 1995}
\begin{abstract}
These lectures present the Fermi-liquid theory of superconductivity, which is applicable to a broad
range of systems that are candidates for non-s wave pairing, {\it e.g.} the heavy fermions, organic
metals and the CuO superconductors. Ginzburg-Landau (GL) theory provides an important link between
experimental properties of non-s wave superconductors and the more general Fermi-liquid theory. The
multiple superconducting phases of UPt$_3$ provide an ideal example of the role that is played by
the GL theory for non-s wave superconductors. The difference between non-s wave superconductivity
and conventional anisotropic superconductivity is illustrated here by the unique effects that
impurities are predicted to have on the properties of non-s wave superconductors.
\end{abstract}
\maketitle

\section{Introduction}

\noindent Historically, ``non-s-wave pairing'' began with the publication
``Generalized Bardeen-Cooper-Schrieffer States and the Proposed
Low-Temperature Phase of $^3$He'' by  Anderson and Morel.\cite{and61}
They considered the possibility  of BCS pairing with non-zero angular
momentum, and studied its physical consequences.  When superfluidity
was discovered in 1972 by Osheroff et al.\cite{osh72} in $^3$He it was
immediately clear that this was not a conventional s-wave BCS
superfluid because  there was more than one superfluid phase.
Increasing  evidence for non-s-wave pairing came from many
 experimental results and within about a year after their discovery the
three superfluid phases of $^3$He were undisputedly identified as
p-wave (as proposed by Anderson and Morel) spin-triplet
superfluids.

The search for ``non-s-wave superconductivity'', the  metallic analog
to superfluidity in $^3$He, was unsuccessful for more than a decade.
In 1979 Steglich \cite{ste79} discovered superconductivity in
CeCu$_2$Si$_2$. This was the first in a new class of heavy-fermion
superconductors, which now include the U-based compounds UBe$_{13}$,
UPt$_3$, URu$_2$Si$_2$, UNi$_2$Al$_3$, and UPd$_2$Al$_3$.  Unusual
temperature dependences of heat capacity, penetration depth, and sound
absorption led to conjectures that these materials were non-s-wave
superconductors.\cite{ott87} Much more experimental information is now
available,\cite{sar92,ste93} and there is consensus that some heavy-fermion
superconductors (if not all of them) show non-s-wave pairing.  The
interest in non-s-wave superconductivity reached new levels  recently
with the reports of several experiments on cuprate superconductors that
supported earlier predictions \cite{bic87} of d-wave pairing.  However,
there is not yet a generally accepted identification of the specific
type of pairing in any of these superconductors.\par

For metals, the popular notion of ``non-s-wave superconductivity''
should not be taken literally. A rigorous classification of
superconductors by the angular momentum of the Cooper pairs (e.g.
s-wave, p-wave, d-wave pairing, etc.) fails in crystalline materials
because angular momentum is not a good quantum number.  There is no
ideally isotropic superconductor in nature, and any superconductor is,
in this  sense, ``non-s-wave''. However, we will use the term
``non-s-wave pairing'' interchangeably with ``unconventional pairing''
or ``reduced symmetry superconductivity'' for superconductors in which
the superconducting state spontaneously breaks one or more symmetries
of the crystalline phase.\par

Liquid $^3$He, heavy-fermion metals, and high-T$_c$ cuprates are
systems of strongly correlated fermions, for which we do not have a
practical microscopic theory. One is forced to rely on phenomenological
theories in order to describe superfluidity or superconductivity in
these systems. This lecture gives an introduction to the two  most
powerful phenomenological theories of correlated fermions with
non-s-wave pairing: the Ginzburg-Landau theory and Fermi liquid
theory.  The Ginzburg-Landau theory is summarized in section [II] where
we discuss the order parameter for non-s-wave pairing, introduce the
basic concepts and notations, and apply the theory to UPt$_3$, the best
studied candidate for non-s-wave pairing in the heavy-fermion
superconductors. Section [III] gives an introduction to the
Fermi-liquid theory of superconductivity. This theory includes the
Ginzburg-Landau theory for $T\sim T_c$, but is more general; it covers
the entire temperature and field range of interest and has proven to be
very powerful in describing the superfluid phases of $^3$He, the
paradigm of a Fermi liquid. Fermi-liquid theory is also very useful
in describing the  heavy-fermion metals  at low temperatures; i.e.
below the ``coherence temperature'' which marks the cross over to the
Fermi-liquid state. Special versions of Fermi-liquid theory (e.g.
nearly antiferromagnetic  Fermi liquid, marginal Fermi liquid,
Eliashberg's strong-coupling model, etc.) seem to be promising steps
toward a theory of high-T$_c$ superconductivity. Finally, in section
[IV] we discuss the effects of impurities on the properties of
non-s-wave superconductors. Impurity scattering leads to several novel
effects in many  non-s-wave superconductors; these examples
demonstrate the predictive power of the Fermi-liquid theory for
unconventional superconductors.\par

This lecture presents selected aspects of the theory of non-s-wave
superconductors, and does not cover the full spectrum of interesting
and promising developments in the theory of unconventional pairing. We
refer the interested reader to the reviews by Lee et al.,\cite{lee86}
Gorkov,\cite{gor87} and Sigrist and Ueda \cite{sig91} for a broader discussion.

\section{Ginzburg-Landau Theory}

\noindent One of the most important phenomenological theories available for
investigating the properties of superconductors is the Ginzburg-Landau
(GL) theory. This theory is applicable to a wide class of materials and
superconducting phenomena, but is limited to temperatures near $T_c$
where the order parameter is small. Here we introduce the order
parameter for non-s-wave superconductors, develop the GL theory from
symmetry considerations and apply the GL theory to UPt$_3$.

The basic quantity that describes all BCS-type superconductors is the
equal-time pair amplitude,
\be\label{pair-amp}
f_{\alpha\beta}(\vec{k}_f)\sim
\left<a_{\vec{k}_f\alpha}a_{-\vec{k}_f\beta}\right>
\,,
\ee
where $\vec{k}_f$ is the momentum of a quasiparticle on the Fermi
surface and $\alpha,\beta$ are the spin labels of the
quasiparticles.\footnote{In the heavy-fermion materials it is generally
assumed that spin-orbit coupling is strong;\cite{and84a,vol85,lee86}
thus, the labels characterizing the quasiparticle states are not
eigenvalues of the spin operator for electrons. Nevertheless, in
zero-field the Kramers degeneracy guarantees that each $\vec{k}$ state
is two-fold degenerate, and thus, may be labeled by a `pseudo-spin'
quantum number $\alpha$, which can take on two possible values. We use
the term `spin' interchangeably with `pseudo-spin' in this article.}
Fermion statistics requires that the pair amplitude obey the
anti-symmetry condition,
$f_{\alpha\beta}(\vec{k}_f)=-f_{\beta\alpha}(-\vec{k}_f)$. Thus, the
most general pairing amplitude can be written in terms of a sum of
spin-singlet and spin-triplet amplitudes,
\be
f_{\alpha\beta}(\vec{k}_f)=f_0(\vec{k}_f)\,(i\sigma_y)_{\alpha\beta}
+\vec{f}(\vec{k}_f)\cdot
(i\vec{\sigma}\sigma_y)_{\alpha\beta}
\,,
\ee
where $f_0(\vec{k}_f)=f_0(-\vec{k}_f)$
($\vec{f}(\vec{k}_f)=-\vec{f}(-\vec{k}_f)$) has even (odd) parity.

The actual realization of superconductivity in a given material will be
described by a pair amplitude of this general form which minimizes the
free energy. The GL theory is formulated in terms of a stationary free
energy functional of the pair amplitude, or order parameter, of
eq.(\ref{pair-amp}). The central assumptions of the GL theory are (i)
that the free energy functional can be expanded in powers of the order
parameter and (ii) that the GL functional has the full symmetry of the
normal state. The GL functional can then be constructed from basic
symmetry considerations.

Essentially all of the candidates for non-s-wave superconductivity,
including the heavy-fermion and cuprate superconductors, have inversion
symmetry. This has an important consequence; the
pairing interaction that drives the superconducting transition
decomposes into even- and odd-parity sectors.\cite{and84a,vol85} Thus,
$f_{\alpha\beta}(\vec{k}_f)$ necessarily has even or odd parity unless
there is a second superconducting instability into a state with
different parity. 
Furthermore, the pairing interaction separates into a sum over
invariant bilinear products of basis functions, ${\cal
Y}_{\Gamma,i}(\vec{k}_f)$, for each irreducible representation $\Gamma$
of the crystal point group, for both even- and odd-parity sectors.
Representative basis functions for the group $D_{6h}$, appropriate for
UPt$_3$ with strong spin-orbit coupling, are given in Table
\ref{tab_basis} (see Ref. (\onlinecite{yip93c}) for a complete
discussion of the allowed basis functions). The general form of the
order parameter is then,
\be\nonumber
f_0(\vec{k}_f)=\sum_{\Gamma}^{even}\,\eta_{i}^{(\Gamma)}\,{\cal
Y}_{\Gamma,i}(\vec{k}_f)
\quad , \quad
\vec{f}(\vec{k}_f)=\sum_{\Gamma}^{odd}\,\eta_{i}^{(\Gamma)}\,\vec{{\cal
Y}}_{\Gamma,i}(\vec{k}_f)
\,.
\ee
There is a single quadratic invariant for each irreducible representation, 
so the leading terms in the GL free energy are of the form,
\be
{\cal F}=\int\,d^3x\,
\sum_{\Gamma}\,\alpha_{\Gamma}(T)\,\sum_{i=1}^{d_{\Gamma}}
|\eta^{(\Gamma)}_{i}|^2 + ...
\,.
\ee
The coefficients $\alpha_{\Gamma}(T)$ are material parameters that
depend on temperature and pressure. Above $T_c$ all the coefficients
$\alpha_{\Gamma}(T)>0$. The instability to the superconducting state is
then the point at which one of the coefficients vanishes, i.e.
$\alpha_{\Gamma^*}(T_c)=0$. Thus, near $T_c$
$\alpha_{\Gamma^*}(T)\simeq\alpha'(T-T_c)$ and  $\alpha_{\Gamma}>0$ for
$\Gamma\ne\Gamma^*$. At $T_c$ the system is unstable to the development
of all the amplitudes $\{\eta_i^{(\Gamma^*)}\}$, however, the higher
order terms in the GL functional which stabilize the system, also
select the ground state order parameter from the manifold of degenerate
states at $T_c$. In most superconductors the instability is in the
even-parity, $A_{1g}$ channel. This is the crystalline analog of
`s-wave superconductivity' in which only gauge symmetry is
spontaneously broken. An instability in any other channel is a
particular realization of non-s-wave superconductivity.

\begin{table}
\begin{tabular}{|c|c|c|c|}
\hline
& Even parity & & Odd parity \\ 
\hline
A$_{1g}$ & 1 &
A$_{1u}$ & $\hat{z}\,k_z$  \\
\hline
A$_{2g}$ & $Im(k_x+ik_y)^6$ &
A$_{2u}$ & $\hat{z}\,k_z Im(k_x+ik_y)^6$  \\
\hline
B$_{1g}$ & $k_z\,Im(k_x+ik_y)^3$ &
B$_{1u}$ & $\hat{z}\,Im(k_x+ik_y)^3$ \\
\hline
B$_{2g}$ & $k_z\,Re(k_x+ik_y)^3$ &
B$_{2u}$ & $\hat{z}\,Re(k_x+ik_y)^3$  \\
\hline
E$_{1g}$ & $k_z\left(\begin{array}{c}k_x \\ k_y \end{array}\right)$ &
E$_{1u}$ & $\hat{z}\left(\begin{array}{c}k_x \\ k_y
 \end{array}\right)$ \\
\hline
E$_{2g}$ & $\left(\begin{array}{c}k_x^2-k_y^2\\2k_xk_y
 \end{array}\right)$ &
E$_{2u}$ & $\hat{z}\,k_z
 \left(\begin{array}{c}k_x^2-k_y^2\\2k_xk_y \end{array}\right)$ \\
\hline
\end{tabular}
\caption{Basis functions for $D_{6h}$}
\label{tab_basis}
\end{table}
\medskip

\subsection*{Application to UPt$_3$}

Considerable evidence in support of non-s-wave superconductivity state
in the heavy-fermion materials has accumulated from specific heat,
upper critical field and various transport measurements, all of which
show anomalous properties compared to those of conventional
superconductors (see Refs. (\onlinecite{ott87,sar92,ste93}) for original  
references). However, the strongest evidence for unconventional
superconductivity comes from the multiple superconducting phases of
UPt$_3$.\cite{mul87,qia87,sch89,bru90,ade90} The important features of
the H-T phase diagram are: (i) There are two zero-field superconducting
phases with a difference in transition temperatures of ${\Delta
T_c}/{T_c}\simeq 0.1$. (ii) A change in slope of the upper critical
field (a `kink' in $H_{c2}^{\perp}$) is observed for
$\vec{H}\perp\hat{c}$, but not for $\vec{H}||\hat{c}$, or at least it
is much smaller. (iii) There are three flux phases, and the phase
transition lines separating the flux phases appear to meet at a
tetracritical point for all orientations of $\vec{H}$ relative to $\hat{c}$.

There are two basic types of models that have been proposed to explain
the phase diagram: (i) theories based on a {\it single} primary order
parameter belonging to a higher dimensional representation of the
symmetry group of the normal state, and (ii) theories based on {\it
two} primary order parameters belonging to different irreducible
representations which are nearly degenerate. Below we construct a
particular GL theory for non-s-wave superconductivity which describes
the superconducting phase diagram of UPt$_3$. We start from the
assumption that the superconducting instability occurs in one of the 2D
representations appropriate to strong spin-orbit coupling, then show
how the presence of weak perturbations can give rise to multiple
superconducting phases.\cite{hes89,mac89} A detailed discussion of this
model for UPt$_3$ is given in Ref. (\onlinecite{sau94b,sau94}).

Consider one of the 2D representations, {\it e.g.} the E$_{2u}$
representation of Table \ref{tab_basis}. Near $T_c$ all other order
parameter amplitudes vanish. The GL order parameter is then a complex
two-component vector, $\vec{\eta}=(\eta_1,\eta_2)$, transforming
according to the relevant 2D representation. The terms in the GL
functional must be invariant under the full symmetry group of point
rotations, time-reversal and gauge transformations. The form of ${\cal
F}$ is governed by the linearly independent invariants that can be
constructed from fourth-order products,
$\sum\,b_{ijkl}\,\eta_i\eta_j\eta_k^*\eta_l^*$, and second-order
gradient terms, $\sum\,\kappa_{ijkl}(D_i\eta_j)(D_k\eta_l)^*$.
For the 2D representations there are only two independent fourth-order
invariants and four independent second-order gradients; the GL free
energy functional has the general form,
\be\label{free_energy}
\begin{array}{l}
\hspace{-1.5em}
\displaystyle{
{\cal F}\left[{\vec{\;\eta},\vec{A\;}}\right] = 
\int\limits d^{3}R \;\Big\lbrace\,\alpha(T)\vec{\,\eta}\cdot\vec{\eta}^{*}
+\beta_{1}\left({\vec{\,\eta}\cdot\vec{\eta}^{*}}
\right)^{2}+\beta_{2}\left\vert{\vec{\eta}\cdot\vec{\eta}}
\right\vert^{2}
+\frac{1}{8\pi}|\vec{\nabla}\times\vec{A}|^2
}
\nonumber\\ 
\hspace{-1.5em}
\qquad\qquad+\kappa_{1}
\left({D_{i}\eta_{j}}\right)\left({D_{i}\eta_{j}}\right)^{*}+\kappa_{2}
\left({D_{i}\eta_{i}}\right)\left({D_{j}\eta_{j}}\right)^{*}+\kappa_{3}
\left({D_{i}\eta_{j}}\right)\left({D_{j}\eta_{i}}\right)^{*}+\kappa_{4}
\left({D_{z}\eta_{j}}\right)\left({D_{z}\eta_{j}}\right)^{*} 
\Big\rbrace 
\,.
\end{array}
\ee
The equilibrium order parameter and current distribution are determined
by the stationarity conditions for variations of ${\cal F}$ with
respect to $\vec{\eta}$ and the vector potential, $\vec{A}$. This
functional provides an important connection between experiment and the
Fermi-liquid theory of superconductivity because the material
parameters, $[\alpha (T), \beta_1, \beta_2, \kappa_1, \kappa_2,
\kappa_3, \kappa_4]$, can be determined from comparison of GL 
theory with experiment, and calculated from Fermi-liquid theory (see below).

There are two possible homogeneous equilibrium states of this GL theory. For $-\beta_1<\beta_2<0$ the equilibrium order
parameter, $\vec{\eta}=\eta_0\hat{x}$ (or any of the six degenerate states obtained by rotation), breaks rotational symmetry in
the basal plane, but preserves time-reversal symmetry. However, for $\beta_2>0$ the order parameter retains
the full rotational symmetry (provided each rotation is combined with
an appropriately chosen gauge transformation), but spontaneously breaks
time-reversal symmetry. The equilibrium state is doubly-degenerate with
an order parameter of the form
$\vec{\eta}_{+}=(\eta_0/\sqrt{2})(\vec{x}+i\vec{y})$ [or
$\vec{\eta}_{-}=\vec{\eta}_{+}^{*}$], where $\eta_0 = (|\alpha | /2
\beta_1)^{1/2}$. The broken time-reversal symmetry of the two
solutions, $\vec{\eta}_{\pm}$, is exhibited by the two possible
orientations of the internal orbital angular momentum, $\vec{l}\sim i
\left({\vec{\eta}\times\vec{\eta}^{\,*}}\right) \sim\pm\,\hat{z}$, or
spontaneous magnetic moment of the Cooper pairs.
 
The case $\beta_2>0$ is relevant for the 2D models of the
double transition of UPt$_3$. However, the 2D theory has only one phase
transition in zero field, and by itself cannot explain the double
transition. The small splitting of the double transition in UPt$_3$
($\Delta T_c/T_c\simeq 0.1$) suggests the presence of a small symmetry
breaking energy scale and an associated lifting of the degeneracy of
the possible superconducting states belonging to the 2D representation.
The second zero-field transition just below $T_c$ in UPt$_3$, as well
as the anomalies observed in the upper and lower critical fields, have
been explained in terms of a weak symmetry breaking field (SBF) that
lowers the crystal symmetry from hexagonal to orthorhombic, and
consequently reduces the 2D E$_2$ (or E$_1$) representation to two 1D
representations with slightly different transition
temperatures.\cite{hes89,mac89} The key point is that right at $T_c$
all states belonging to the 2D representation are degenerate, thus any
SBF that couples second-order in $\vec{\eta}$ and prefers a particular
state will dominate very near $T_c$. At lower temperatures the SBF
energy scale, $\Delta T_c$, is a small perturbation compared to the
fourth order terms in the fully developed superconducting state and one
recovers the results of the GL theory for the 2D representation, albeit
with small perturbations to the order parameter.
 
In UPt$_3$ there appears to be a natural candidate for a
SBF,\cite{joy88} the AFM order in the basal plane reported by Aeppli,
et al.\cite{aep88} In this case a
contribution to the GL functional corresponding to the coupling of the
AFM order parameter to the superconducting order parameter is included;
${\cal F_{{\rm SBF}}}\left[{\vec{\;\eta}}\right]
=\varepsilon\,M_s^2\,\int\limits d^{3}R\,\left(|\eta_1|^2 -
|\eta_2|^2\right)$, where $\vec{M}_s$ is the AFM order parameter and
the coupling parameter $\varepsilon M_s^2$ determines the magnitude of
the splitting of the superconducting transition. The analysis of this
GL theory, including the SBF, is given in Ref. (\onlinecite{hes89}); the
main results are:
\begin{enumerate}
\item A double transition occurs only if $\beta_2>0$.
The splitting of the transition temperature is $\Delta T_c\propto
M_s^2$.
\item The relative magnitudes of the two heat capacity anomalies,
$\Delta C_*/T_{c*} > \Delta C/T_c$, are consistent with
the stability condition $\beta_2>0$ required by the double transition in the 2D model.
\item The low temperature phase ($T<T_{c*}$) has broken time-reversal
symmetry, and is doubly degenerate: $\vec{\eta}_{\pm}\sim (a(T),\pm
i\,b(T))$, reflecting the two orientations of the internal angular
momentum of the ground state.
\item The upper critical field exhibits a change in slope, a `kink' at
high temperature for $\vec{H}\perp\hat{c}$, but not for
$\vec{H}||\hat{c}$. The kink in $H_{c2}^{\perp}$ is isotropic in the
basal plane provided the in-plane magnetic anisotropy energy is weak
compared to the Zeeman energy acting on $\vec{M}_s$, in which case
$\vec{M}_s$ rotates to maintain $\vec{M}_s\perp\vec{H}$. Recent
magnetoresistance experiments support this
interpretation of weak magnetic anisotropy energy in the basal
plane.\cite{vor92}
\item Kinks in $H_{c1}$, for all field orientations, are predicted to
occur at the second zero-field transition temperature, $T_{c*}$.
This was confirmed by several groups.\cite{shi89,vin91,kne92,vor92}
The increase in $H_{c1}\propto \eta^2$ at $T_{c*}$ is also a strong
indication of the onset of a second superconducting order parameter.
\item Additional evidence for a SBF model of the 
double transition comes from pressure studies of 
the superconducting and AFM phase transitions. Heat capacity measurements 
by Trappmann, et al.\cite{tra91} show that both zero-field transitions are suppressed 
under hydrostatic pressure, and that the double transition disappears 
at $p_{*}\simeq 4\,kbar$. Neutron scattering experiments reported by 
Hayden, et al.\cite{hay92} show that AFM order disappears on the same pressure 
scale, at $p_c\simeq 3.2\,kbar$.
\end{enumerate}

The phase diagram determined by ultrasound velocity measurements
indicates that the phase boundary lines meet at a tetracritical point
for both $\vec{H}||\hat{c}$ and $\vec{H}\perp\hat{c}$. This has been
argued to contradict the GL theory based on a 2D order
parameter.\cite{luk91,mac91,che93} The difficulty arises from gradient
terms of the form, $\kappa_{23}\left[(D_x\eta_1)(D_y\eta_2)^* +
(D_x\eta_2)(D_y\eta_1)^* +c.c.\right]$, that couple the two components
of the order parameter.  These terms lead to `level repulsion' effects
in the linearized GL differential equations which prevent the crossing
of two $H_{c2}(T)$ curves, corresponding to different superconducting
phases. This feature of the 2D model has spawned alternative theories,
designed specifically to eliminate the `level repulsion'
effect,\cite{mac91,che93} and recently to a more specific version of
the 2D model coupled to a SBF.\cite{sau94b} We discuss this latter model
below and direct readers to Refs.
(\onlinecite{luk91,mac91,che93,sau94}) for discussions of alternative
GL models for UPt$_3$.

Although the GL functionals are formally the same for any of the 2D
orbital representations, the predictions for the material parameters
from Fermi-liquid theory differ substantially depending on the symmetry
properties of the Cooper pair basis functions. For example, the
interpretation of the unusual anisotropy of $H_{c2}$ in UPt$_3$ in
terms of anisotropic Pauli limiting requires an odd-parity,
spin-triplet representation with the $\vec{d}$-vector parallel to the
$\hat{c}$ direction.\cite{cho91} This restricts one to the E$_{2u}$ or
E$_{1u}$ basis functions among the four possible 2D representations.
The case for an odd-parity order parameter with $\vec{d}||\hat{c}$ is
discussed in detail in Refs. (\onlinecite{cho91,sau94b,sau94}).

There are several other important predictions from Fermi-liquid theory
for the 2D representations.  For any of the four 2D representations,
and independent of the geometry of the Fermi surface or the presence of
non-magnetic, s-wave impurity scattering, the fourth-order free energy
coefficients have the ratio,
$\frac{\beta_2}{\beta_1}=\frac{1}{2}$.\cite{sau94a} This ensures that
the coupling to the SBF will produce a double transition in zero field
for any of the 2D orbital representations.  Differences between the 2D
models appear for the in-plane gradient terms in the GL functional.
These gradient coefficients can be calculated from quasiclassical
theory; in the clean limit the results are\cite{sau94a}
\be\label{kappas}
\kappa_1= \kappa_{0}\,
\left<{\cal Y}_1(\vec{k}_f)\,v_{fy}\,v_{fy}\,{\cal Y}_1(\vec{k}_f)\right>
\quad , \quad
\kappa_2= \kappa_3=
\kappa_{0}\,
\left<{\cal Y}_1(\vec{k}_f)\,v_{fx}\,v_{fy}\,{\cal Y}_2(\vec{k}_f)\right>
\,,
\ee
where ${\cal Y}_i(\vec{k}_f)$ are the basis functions,
$\kappa_{0}=\frac{7\zeta(3)}{16\pi^2 T_c^2}\,N_f$, and $N_f$ is the
density of states at the Fermi level. There are important differences
between the E$_{1}$ and E$_{2}$ representations when we evaluate
these averages. Using the basis functions in Table \ref{tab_basis} and
a Fermi surface with weak hexagonal anisotropy, the E$_{1}$ model gives
$\kappa_2=\kappa_3\simeq\kappa_1$, while for the E$_{2}$ model,
$\kappa_2=\kappa_3\ll\kappa_1\sim N_f(v_f^{\perp}/\pi T_c)^2$.
In fact, the three in-plane coefficients are identical for E$_{1u}$ in
the limit where the in-plane hexagonal anisotropy of the Fermi surface
vanishes. In contrast, the coefficients $\kappa_2$ and $\kappa_3$ for
the E$_{2u}$ model both vanish when the hexagonal anisotropy of the
Fermi surface is neglected. This latter result follows directly from
the approximation of a cylindrically symmetric Fermi surface and Fermi
velocity, $\vec{v}_f = v_f^{\perp}(\hat{k}_x\,\vec{x} +
\hat{k}_y\,\vec{y}) + v_f^{||}\hat{k}_z\vec{z}$, and the higher angular
momentum components of the E$_{2}$ basis functions,
\be
\kappa_2(E_{2u})
\propto\left<\hat{k}_z(\hat{k}_x^2-\hat{k}_y^2)v_{fx} v_{fy} (2\hat{k}_x \hat{k}_y)\hat{k}_z\right>\equiv 0
\,.
\ee
If we have weak hexagonal anisotropy of the
Fermi surface then $\kappa_{23}\ll\kappa_1$ only for E$_{2}$.  Thus,
there is a natural explanation for the absence (or at least the
smallness) of the `level repulsion' terms in the orbital 2D model if
there is weak hexagonal anisotropy in the basal
plane, but we are required to select the E$_{2}$ representation. There
is support for the assumption of weak hexagonal anisotropy; if the
hexagonal anisotropy of $\vec{v}_f$ were significant it should be
observable at low temperature as an in-plane anisotropy of
$H_{c2}^{\perp}(T)$. The angular dependence of $H_{c2}^{\perp}$ was
investigated, but no in-plane anisotropy was observed.\cite{shi86a}

In order to account for the discontinuities in the slopes of the
transition lines near the tetracritical point an additional
ingredient in the GL theory for the E$_{2u}$ model is needed that is
not contained in the theory of Hess, {\it et al.}\cite{hes89}
For E$_{2u}$ with $\kappa_{23}=0$ the gradient energy reduces to
${\cal F}_{grad}=\int\,d^3x \left\{
\kappa_1\right(|\vec{D}_{\perp}\eta_1|^2 + |\vec{D}_{\perp}\eta_2|^2\left) +
\kappa_4\right(|D_z\eta_1|^2 + |D_z\eta_2|^2\left)
\right\}$.
Because both order parameter components appear with the same
coefficients there is no crossing of different $H_{c2}(T)$ curves
corresponding to different eigenfunctions, and therefore no
tetracritical point. However, the slopes of the transition lines near
the tetracritical point suggest that the difference in the gradient
energies associated with the two components of the order parameter are
finite, but small, i.e. $|\Delta\kappa/2\kappa_1|\lesssim
0.2$. This suggests that the SBF may be responsible for a
splitting in the gradient coefficients as well as the transition
temperature.

In the model of Hess, et al.\cite{hes89} the coupling to the SBF was
included through second order in both the superconducting order
parameter, $\vec{\eta}$, and the AFM order parameter, $\vec{M}_s$, but
only for the homogeneous terms in the free energy. However, there is
also a contribution to the gradient energy that is second-order in
$M_s$. These terms are as essential for describing a double transition
as a function of field, as the homogeneous term is for the double
transition in zero field. The relevant invariants are of the form
$M_s^2(|\vec{D}_{\perp}\eta_1|^2 - |\vec{D}_{\perp}\eta_2|^2)$ and
$M_s^2(|D_z\eta_1|^2 - |D_z\eta_2|^2)$, and lead to a gradient energy
\be
{\cal F}_{grad}=\int\,d^3x \left\{
(\kappa_1^{+}|\vec{D}_{\perp}\eta_1|^2 + 
\kappa_1^{-}|\vec{D}_{\perp}\eta_2|^2)
+
(\kappa_4^{+}|D_z\eta_1|^2 + 
\kappa_4^{-}|D_z\eta_2|^2)
\right\}
\,,
\ee
where
$\kappa_1^{\pm} = \kappa_1 (1 \pm \epsilon_{\perp} M_s^2)$ and
$\kappa_4^{\pm} = \kappa_4(1 \pm \epsilon_{||} M_s^2)$.
Dimensional analysis implies that the coupling coefficients,
$\epsilon$, $\epsilon_{\perp}$, $\epsilon_{||}$, for the homogeneous
term, the in-plane gradient energies and the $\hat{c}$-axis gradient
energies are formally the same order of magnitude, in which case we conclude
that the splittings in the gradient coefficients are relatively small,
\be
\left|\frac{\kappa_{1,4}^{+}-\kappa_{1,4}^{-}}{2\kappa_{1,4}}\right|=
\left|\epsilon_{\perp,||}\,M_s^2\right|\sim\left|\frac{\Delta T_c}{T_c}\right|
\,.
\ee
Thus, in this GL theory the SBF is essential for producing an apparent
tetracritical point, and at this semi-quantitative level, can account
for the magnitudes of the slopes near the tetracritical
point.\cite{sau94b} As this model for UPt$_3$ illustrates, the GL theory
provides a central link between experiments, in this case the phase
diagram, and the more microscopic Fermi-liquid theory which we now
discuss.

\section{Fermi-Liquid Theory}

\noindent Conduction electrons in metals interact strongly with each other and
with the lattice. These interactions lead to correlations among the
electrons, and we have to view conduction electrons, in general, as a
system of correlated Fermions. On the other hand, the model of
non-correlated electrons occupying single-particle states in conduction
bands describes most properties of traditional metals very well. This
simple behaviour of an intrinsically complicated system was explained
by Landau in a series of papers in which he established his
``Fermi-liquid theory.'' Landau argued that interacting Fermions can
form a ``Fermi-liquid state,'' which for many properties has the
appearance of a non-correlated state. The reason, according to Landau,
is that the physical properties of a Fermi liquid at low temperatures
are dominated by low-lying excitations (quasiparticles) which are
complex objects but have basic features (e.g. charge, spin, fermion
number) in common with non-interacting electrons. Landau showed that an
ensemble of quasiparticles is described by a classical
distribution function $g(\vec k_f,\vec R;\epsilon,t)$,\footnote{We
use the ``energy representation'' in which the traditional variables
$\vec k$, $\vec R$, $t$ of the distribution function are replaced by
the equivalent set $\vec k_f$, $\epsilon$, $\vec R$, $t$. This amounts
to a transformation to new coordinates in momentum space. The
3-dimensional momentum variable ($\vec k$) is replaced by the
2-dimensional momentum variable $\vec k_f$, which is the point on the
Fermi surface nearest to $\vec k$, and the 1-dimensional energy
variable $\epsilon=E(\vec k,\vec R, t)-E_f$. The energy representation
of the Boltzmann-Landau equation has a wide range of validity. It also
describes overdamped excitations whose lifetime is comparable to or
shorter than its oscillation period $\hbar/\epsilon$. For a review on
transport theory in energy representation see Ref. (\onlinecite{ram86}).} and that
this distribution function obeys a classical transport equation
(Boltzmann-Landau transport equation).

Derivations of the Boltzmann-Landau transport equation from first
principles \cite {eli62,pra64} use many-body Green's function techniques, and lead to
explicit expressions for the various terms of the transport equation in
terms of self-energies. The self-energies describe the effects of
electron-electron and electron-phonon interactions as well as impurity
scattering. The complete set of diagrams for the leading order
self-energies are shown in Fig. (I); the full circles are
vertices representing quasiparticle-quasiparticle,
quasiparticle-phonon, and quasiparticle-impurity interactions. We
follow Landau and consider these vertices together with the Fermi
surface properties (e.g. shape of the Fermi surface and Fermi
velocities) as phenomenological parameters of the Fermi-liquid model,
but in principle they can be obtained from the full many-body theory.
\par

The present notion of a normal Fermi liquid is somewhat more general
than standard definitions. For the conduction electrons of a metal to
form a Fermi liquid we require that \begin{itemize} \item the dominant
charge-carrying excitations have the signatures of electrons or holes;
they are fermions with charge $\pm e$, spin 1/2, {\it large}
velocities $\vec v_f$ and {\it large} momenta $\vec k_f$ near a Fermi
surface.\footnote{The term {\it large} means that the corresponding energies
($k_f^2/2m^{\ast} \sim m^{\ast}v_f^2/2 \sim E_f$) are much greater than
typical excitation energies $\epsilon\sim k_BT_c$.}
\item The physical properties of the ensemble of these excitations can
be described by a Boltzmann-Landau transport equation. \end{itemize}
In the following we call the electronic quasiparticles simply
``electrons'', but emphasize that the physical properties of such an
electron are, in general, significantly affected by many-body effects.
\par

Landau's Fermi-liquid model predicts a number of universal results for
temperature and magnetic field dependences of thermodynamic and
transport properties at very low temperatures. The universal laws can
be used as experimental signatures for Fermi-liquid behaviour,
provided the low-temperature limit is reached before the
superconducting transition. A detailed discussion of these normal-state
properties can be found in various textbooks and review articles.
\cite{baym91} The systems with a potential interest for non-s-wave
superconductivity, i.e. heavy-fermion superconductors and
high-T$_c$ superconductors, often have not yet reached their ideal
low-temperature limit when superconductivity sets in. In such systems
the investigation of Fermi-liquid behaviour requires an analysis of
measurements in both the normal and superconducting phases. We
focus on the Fermi-liquid model of the superconducting phases.

\begin{figure}
\includegraphics[width=0.75\columnwidth]{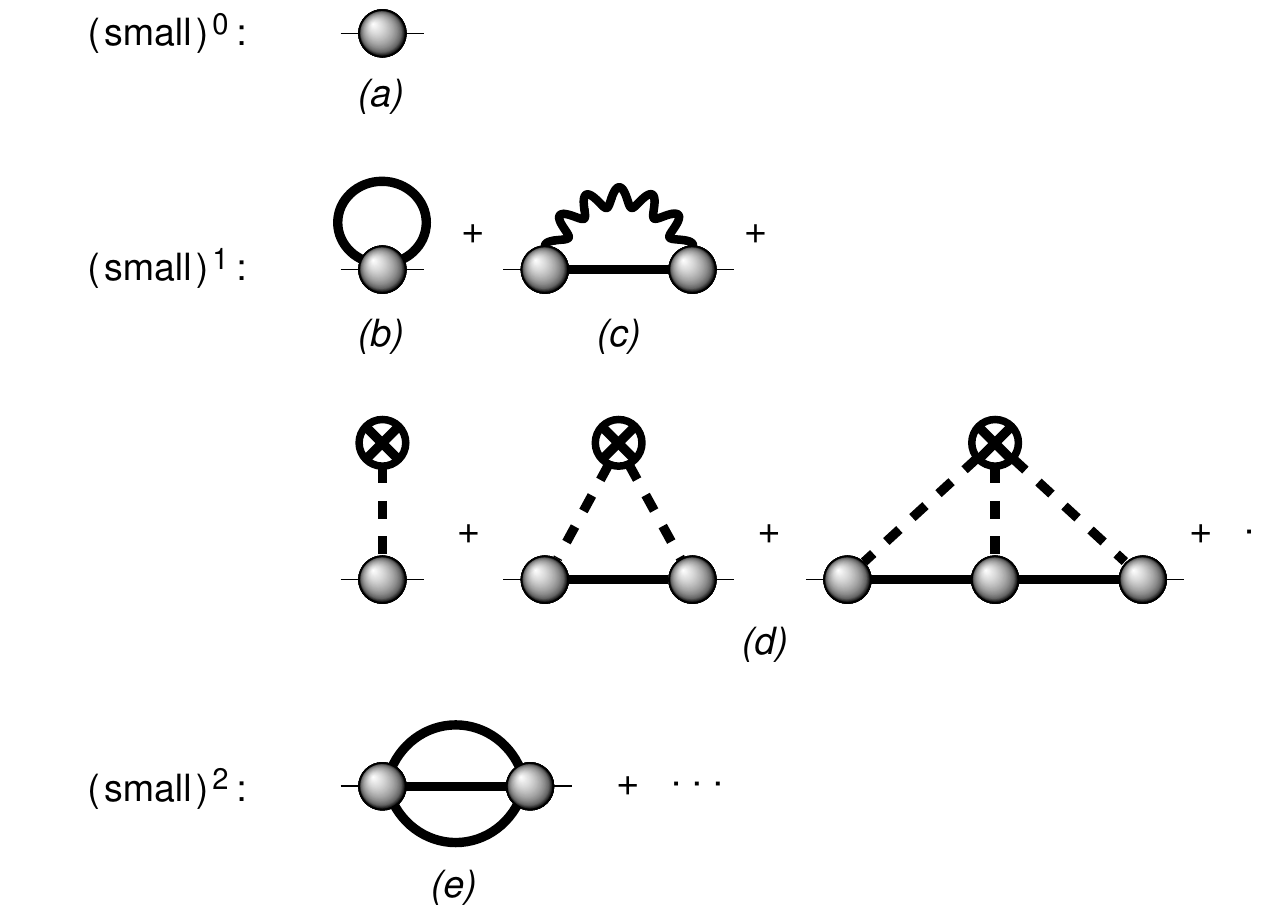}
\caption{
Leading order self-energy diagrams of Fermi-liquid theory. The vertices (shaded
circles) represent the sum of all high-energy processes and give rise
to interactions between the quasiparticles (smooth propagator lines),
phonons (wiggly propagator lines) and impurities (dashed lines). The
order in the parameter 'small'  is indicated for each diagram.}
\end{figure}

\subsection*{Superconducting Fermi Liquids} 
 
It took more than 10 years after the breakthrough in the theory of
superconductivity by BCS to establish a complete Fermi-liquid theory of
superconductivity. Earlier theories of superconductivity were
formulated in terms of Bogolyubov's equations, Gorkov's equations, or
other fully quantum mechanical schemes. They lacked the quasiclassical
aspects of Landau's Fermi-liquid theory. The first complete
quasiclassical theory of superconductivity was formulated in a series
of publications by Eilenberger,\cite{eil68} Larkin and Ovchinnikov,
\cite{lar69,lar75} and Eliashberg.\cite{eli72} It is
presented and discussed in several review articles.~\cite{eck81,ser83,lar86,ram86}
The quasiclassical theory allows one
to calculate all superconducting phenomena of interest, including
transition temperatures, excitation spectra, Josephson effects,
vortex structures, the electromagnetic response, etc. In this theory
the dynamical degrees of freedom of electronic quasiparticles are
described partly by classical statistical mechanics, and partly by
quantum statistics. The classical degrees of freedom are the motion in
$\vec k$-$\vec R$ phase space; i.e. quasiparticles move along
classical trajectories. Of special importance for understanding
superconducting phenomena are the quantum-mechanical ``internal degrees
of freedom'', the {\sl spin} and the {\sl particle-hole}
degrees of freedom of an electron. Quantum coherence between particle
excitations and hole excitations is a key feature of the BCS theory of
superconductivity and the origin of all non-classical effects in
superconductors (e.g. supercurrents, coherence factors in transition
amplitudes, Andreev reflection, etc.). Particle-hole coherence is taken
care of in the quasiclassical theory by grouping particle excitations
(occupied one-electron states with energy above the Fermi energy
($\epsilon >0$)) and hole excitations (empty one-electron states with
$\epsilon <0$) into a doublet. The spin and particle-hole doublets
form a 4-dimensional Hilbert space of internal degrees of freedom for
quasiparticle excitations. In quantum statistical mechanics the
internal state of quasiparticle excitations is described by a $4\times
4$ density matrix. Hence, the distribution function of the
quasiclassical theory is a $4\times4$-matrix whose elements are
functions of $\vec k_f$, $\vec R$, $\epsilon$, and $t$.\par

The central object of the quasiclassical theory is the ``Keldysh
quasiclassical propagator'' $\hat g^K(\vec k_f,\vec R;\epsilon,t)$. It
is a $4\times 4$-matrix which generalizes the classical
Boltzmann-Landau distribution function to the superconducting state,
and carries the information on both the classical and quantum degrees
of freedom. We denote $4\times 4$-matrices by a ``hat''. In addition,
we use standard notation and introduce a superscript on the
quasiclassical propagators to identify their microscopic origin. Thus,
$\hat g^K$ denotes the Keldysh propagator which is obtained by
$\xi$-integration \cite{ser83} from the microscopic Keldysh Green's
function. The 16 matrix elements of $\hat g^K$ can be expressed in
terms of 4 spin-scalars ($g^K$, $\underline{g}^K$, $f^K$,
$\underline{f}^K$) and 4 spin-vectors ($\vec g^K$,
$\vec{\underline{g}}^K$, $\vec f^K$, 
$\underline{\vec f}^{\raisebox{-1ex}{$\scriptstyle K$}}$):
\begin{equation}\label{matrx} 
\hat g^K=
\left(
\begin{array}{cc}
g^K+\vec g^K
\cdot\vec\sigma&\left(f^K
+\vec f^K\cdot\vec\sigma\right)i\sigma_y\\
i\sigma_y\left(\underline{f}^K+
\underline{\vec f}^{\raisebox{-1ex}{$\scriptstyle K$}}\cdot\vec\sigma
\right)&\underline{g}^K
-\sigma_y\underline{\vec g}^K
\cdot\vec \sigma\sigma_y
\end{array}\right)\ . 
\end{equation}
Retarded and advanced quasiclassical propagators, 
$\hat{g}^{R,A}(\vec k_f,\vec R,;\epsilon,t)$, are defined analogously 
by $\xi$-integration
of the retarded and advanced microscopic Green's functions.\cite{ser83} 
The diagonal components of (\ref{matrx}) are directly
related to scalar distribution functions for particles, $n_p(\vec
k_f,\vec R;\epsilon,t)$, and holes, $n_h(\vec k_f,\vec
R;\epsilon,t)$:\footnote{ $n_{p(h)}(\vec k_f,\vec R;\epsilon,t)d^2k_fd^3R\,
d\epsilon$ is the number of particle excitations
(hole excitations) with momentum $\vec k_f$, position $\vec R$, and
energy $\epsilon$ in the phase space element $d^2k_fd^3Rd\epsilon$.}
\begin{equation}\label{dis1}
n_p=
\frac{-i}{(2\pi)^4\mid\vec v_f\mid}
\left(g^K -(g^R-g^A)\right), \ \ 
n_h=\frac{+i}{(2\pi)^4\mid\vec v_f\mid}
\left(
\underline g^K -(\underline g^R -\underline g^A) \right) 
\end{equation}
Similarly, $\vec g^K$ and $\vec{\underline{g}}^K$ are related to
spin-distribution functions. The off-diagonal components $f^K( \vec
k_f,\vec R;\epsilon,t)$, $\vec f^K( \vec k_f, \vec R;\epsilon,t)$ in
(\ref{matrx}) can be interpreted as pair amplitudes.
\footnote{The amplitudes $\mbox{\underline{f}}^{K}$  and 
$\vec{\mbox{\underline{f}}}^{\,K}$
are redundant since they are related to $f^K$ and $\vec{f}^K$ by fundamental
symmetry relations.\cite{ser83}} 
They carry the information on
particle-hole coherence. Non-zero pair amplitudes indicate
particle-hole coherence and thus superconductivity. More specifically,
a non-vanishing scalar pair amplitude, $f^K$, (vector pair amplitude
$\vec{f}^K$) describes even parity, spin-singlet pairing (odd parity,
spin-triplet pairing).\par

Measurable quantities such as the charge density $n(\vec R,t)$, current
density $\vec j(\vec R,t)$, etc. can be calculated from the diagonal
components of the quasiclassical propagator, $\hat g^K$, or,
equivalently, from the distribution functions $n_p$ and $n_h$. If we
ignore the Landau parameters (contained in diagram Ib), then
\begin{eqnarray}
n(\vec R,t)
&=& 
\displaystyle{
n_0+ 2e^2N_f\Phi(\vec R,t)+
2e
\int{d^2k_f\over (2\pi)^3\mid \vec v_f\mid}
\int_{-\infty}^{\infty}{d\epsilon\over 4\pi i}
g^K(\vec k_f,\vec R;\epsilon,t)
}
\nonumber\\
&=& 
\displaystyle{
n_0+ 2e^2N_f\Phi(\vec R,t)+
e
\int{d^2k_f\over (2\pi)^3\mid \vec v_f\mid}
\int_0^{\infty}{d\epsilon}
\left(n_p -
n_h +{(g^R-g^A)+(\underline g^R- \underline g^A)\over (2\pi i)}\right)
}
\,,
\label{nden} 
\\
\vec j(\vec R,t)
&=& 
2e
\int{d^2k_f\over (2\pi)^3\mid \vec v_f\mid}
\int_{-\infty}^{\infty}{d\epsilon\over 4\pi i}
\,\vec v_f\,g^K(\vec k_f,\vec R;\epsilon,t)
\nonumber\\
&=&
e
\int{d^2k_f\over (2\pi)^3\mid \vec v_f\mid}
\int_0^{\infty}{d\epsilon}\,\vec v_f
\left(n_p + n_h+{(g^R-g^A)-(\underline g^R- \underline g^A)\over 
(2\pi i)}\right)\
\,.
\label{ncur}
\end{eqnarray}
In eq. (\ref{nden}), $n_0+ 2e^2N_f\Phi(\vec R,t)$ is the local
equilibrium charge density of electrons in the presence of an
electro-chemical potential $\Phi$. The terms involving $n_p$, $n_h$ in
eqs.(\ref{nden}, \ref{ncur}) represent the contributions of thermally
excited particles and holes, whereas the terms involving $g^{R,A}$,
$\underline g^{R,A}$ are contributions from the superconducting
condensate. Note that the hole-distribution function enters the
density and the current with different signs, because electrons and
holes with the same momentum $\vec k_f$ have opposite charge but carry
the same current.\par

The central equation of the quasiclassical theory is the quasiclassical
transport equation for $\hat g^K$, and is the generalization of the
Landau-Boltzmann equation to the superconducting state. It is a set of
coupled partial differential equations of first order in the spatial
derivatives and, in general, of infinite order in the time derivative.
In a compact notation, the quasiclassical transport equation reads
\begin{equation}\label{qcltr} 
\hspace{-1.5em}\left(\epsilon\hat\tau_3-\hat v-\hat
\sigma^R\right)\otimes
\hat g^K-
\hat g^K\otimes
\left(\epsilon\hat\tau_3-\hat v-\hat
\sigma^A\right)-
\hat\sigma^K\otimes
\hat g^A+ \hat g^R
\otimes 
\hat\sigma^K+i\vec v_f\cdot\vec\nabla 
\hat g^K=0,
\end{equation}
where the $\otimes$-product stands for the usual $4\times 4$-matrix
product and a product in the energy-time variables defined by $\hat
a\otimes\hat b(...;\epsilon,t)=
exp\left(i(\partial_{\epsilon}^a\partial_{t}^b-
\partial_{t}^a\partial_{\epsilon}^b)/2\right)\hat a(...;\epsilon,t)
\hat b(...;\epsilon,t)$. The superscripts $a$ ($b$) on the partial
derivatives indicate derivatives with respect to the arguments of $\hat
a$ ($\hat b$). A determination of $\hat g^K$ from the transport eq.
(\ref{qcltr}) requires knowledge on the external potentials $\hat
v(\vec k_f,\vec R,t)$, the advanced, retarded and Keldysh self-energies
$\hat \sigma^{R,A,K}( \vec k_f,\vec R;\epsilon,t)$, and the advanced
and retarded quasiclassical propagators $\hat g^A(\vec k_f,\vec
R;\epsilon,t)$ and $\hat g^R(\vec k_f,\vec R;\epsilon,t)$. These
propagators are auxiliary quantities which in general have no direct
physical interpretation, except in the adiabatic limit \cite{ser83}
where they determine the local quasiparticle density of states
$N(\vec k_f,\vec R;\epsilon,t)$ via the relation $N(\vec k_f,\vec
R;\epsilon,t)=-N_f\Im m\left(g^R(\vec k_f,\vec
R;\epsilon,t)\right)/\pi$. Finally, in equilibrium $\hat g^K$ is given
in terms of $\hat g^{R,A}$ by $\hat g^{K}= \left(\hat g^{R}- \hat
g^{A}\right)\tanh({\epsilon/ 2T})$.\par

The solutions to the quasiclassical equations for $\hat g^{R,A}$,
\begin{equation}\label{qcltrRA} 
\left[\epsilon\hat\tau_3-\hat v-\hat
\sigma^{R,A},
\hat g^{R,A}\right]_{\otimes} 
+i\vec v_f\cdot\vec\nabla \hat g^{R,A}=0\ ,
\end{equation}
are necessary inputs to the transport eq. (\ref{qcltr}), and the set
of physically relevant solutions of eqs. (\ref{qcltr}, \ref{qcltrRA})
must satisfy the normalization conditions,
\begin{equation}\label{norm1} 
\hat g^{R,A}\otimes 
\hat g^{R,A}=-\pi^2\hat 1\ , \ \ 
\hat g^{R}\otimes 
\hat g^{K}+ 
\hat g^{K}\otimes 
\hat g^{A}=0\ .
\end{equation} 

The quasiclassical equations (\ref{qcltr}, \ref{qcltrRA}) need to be
complemented by self-consistency equations for the the quasiclassical
self-energies $\hat\sigma^{R,A,K}$. They give $\hat\sigma^{R,A}$ in
terms of $\hat g^{R,A}$, and $\hat\sigma^{K}$ in terms of $\hat g^{R}$,
$\hat g^{A}$ and $\hat g^{K}$. The self-consistency equations are
shown, in a diagrammatic notation, in Fig. (I). For example,
these equations include the "gap equation", which is the
self-consistency equation for the off-diagonal self-energy. Explicit
forms of the quasiclassical self-consistency equations can be found in
the reviews.~\cite{eck81,ser83,lar86,ram86}

This completes our list of equations of the theory of Fermi-liquid
superconductivity. The quasiclassical equations, normalization
conditions, and self-consistency equations (Fig. (I)) form a set of
non-linear integro-differential equations. They are exact to leading
orders in the expansion parameters of Fermi-liquid theory (e.g.
$k_BT/E_f$, $1/k_f\xi_0$, $1/k_f\ell$, $\omega_D/E_f$,  ...), and cover equilibrium
phenomena as well as non-equilibrium superconductivity.

\subsection*{Non-S-Wave Pairing} 

The quasiclassical theory is especially well suited for studying
non-s-wave superconductors. Many calculational steps of the theory are
identical for s-wave and non-s-wave superconductors, and many s-wave
results can be simply generalized to anisotropic, non-s-wave wave
superconductors. The reason is that a quasiparticle at point $\vec R$
moving with momentum $\vec k_{f}$ senses only the gap function
$\Delta(\vec k_{f},\vec R)$. Since it does not sample the full momentum
dependence of the gap function it cannot distinguish an anisotropic gap
from an isotropic one. For example, the density of states $N(\vec
k_f;\epsilon)$ of a homogeneous superconductor in equilibrium has the
traditional BCS form, $N(\epsilon,\vec k_f)=N_f\Re
e(\epsilon/\sqrt{\epsilon^2-\mid\Delta\mid^2})$,
but with the isotropic BCS gap $\mid\Delta\mid$ replaced by an
anisotropic local gap $\mid \Delta(\vec k_f)\mid$ for singlet pairing
or $\mid\vec\Delta(\vec k_f)\mid$ for unitary triplet pairing.\par

Several measurable properties of a non-s-wave superconductor can be
calculated in a rather simple way by using standard s-wave formulas
locally at any point $\vec k_f$ on the Fermi surface, and then
averaging over the Fermi surface. Typical examples for such averaged
quantities are the free energy of a superconductor (and related
quantities such as the critical field $H_c(T)$ and the specific heat
$c_s(T)$), the stripped penetration depth tensor,
$\lambda_{ij}$,\footnote{The measured penetration depth is affected
(dressed) by quasiparticle interactions.\cite{leg65} We follow the
terminology of superfluid $^3$He and use the adjective ``stripped'' for
properties of non-interacting quasiparticles.}
\begin{eqnarray}
&
\displaystyle{
{H_c^2(T)\over 8\pi}=N_f\left\langle {\mid\Delta(\vec
k_f)\mid^2}
\Bigg(\ln(T_c/T)-\pi T\sum_{n=-\infty}^{+
\infty}{\mid\Delta(\vec k_f)
\mid^2\over \mid\epsilon_n\mid\left(\mid\epsilon_n\mid
+\sqrt{\epsilon_n^2+\mid\Delta(\vec
k_f)\mid^2}\right)^2}\Bigg)\right\rangle
}
\label{hc}
\\
&
\displaystyle{
\left(\lambda^{-2}\right)_{ij}=
N_f{4\pi e^2\over c^2}\pi T
\sum_{n=-\infty}^{+\infty}\left\langle{
v_{fi}(\vec k_f)v_{fj}(\vec k_f)\mid\Delta(\vec
k_f)\mid^2\over \left[\epsilon_n^2+\mid\Delta(\vec k_f)\mid^2\right]^{3/2}}
\right\rangle
}
\,,
\label{penet}
\end{eqnarray}
where $\epsilon_n=(2n+1)\pi T$ are the Matsubara energies and the Fermi surface average is, 
\begin{equation}\label{avr} 
\Big\langle \ ... \ \Big\rangle ={1\over N_f}\int 
{d^2k_f\over(2\pi)^3\mid \vec v_f\mid}\ ...\ \ . 
\end{equation}
Eq. (\ref{penet}) is the clean limit of a general formula derived in 
Refs.~(\onlinecite{cho88,cho89b}).\par

For these averaged properties there is no fundamental difference between anisotropic conventional
superconductors and genuine unconventional superconductors. One significant feature are here the
zeros of the energy gap on the Fermi surface. These zeros (points or lines of zeros) determine the
phase space for quasiparticle excitations, and lead to characteristic temperature dependences for
$T\rightarrow 0$ which might be used to identify unconventional pairing. Zeros of the gap in
unconventional superconductors are often a direct consequence of the symmetry of the order
parameter, and thus robust features. They lead to power laws in the $T$-dependence of Fermi-surface
averaged quantities, whereas superconductors with a finite minimum gap on the Fermi surface show an
exponential $T$-dependence at low temperatures. The various power laws predicted by theory for
unconventional superconductors are discussed in the reviews. \cite{gor87,sig91}\par

The most striking differences between s-wave and non-s-wave superconductors, which are intimately
connected with unconventional superconducting order parameters, are (i) complex phase diagrams due
to the existence of a variety of different superconducting phases (section II), (ii) new types of
vortices and other defects in the order parameter field,\cite{tok90,tok90a,pal90,bar91,pal92,mel92}
(iii) new collective modes which affect the electromagnetic response and other dynamical response
functions,\cite{hir89,yip92} (iv) a significant effect on the order parameter due to impurities
(section IV), surfaces and interfaces,\cite{amb74,kur87} (v) anomalous Josephson
effects.\cite{ges85,mil88,yip93,yip94} All these anomalous collective effects have been observed
in superfluid $^3$He, which is the best studied unconventional ``superconductor''. By contrast, only
a few of these indicators for non-s-wave superconductivity have been observed in superconducting
metals. For example, the $H$-$T$ phase diagram of UPt$_3$ shows a variety of superconducting phases
and provides evidence for unconventional superconducting order parameter (see section II). And in
high-T$_c$ superconductors (YBCO) anomalous Josephson effects have recently been
reported.\cite{van94}
 
Collective effects are different in origin than the ``averaged effects'' discussed above. They probe
the coherence of quasiparticles with different momenta $\vec k_f$, i.e., quasiparticles moving along
different classical trajectories. The coupling of quasiparticles with different momenta is a
consequence of quasiparticle interactions and self-energy terms in the quasiclassical equations. The
self-consistency equations for the self-energies are integrals over the Fermi surface which sample,
in general, the full anisotropy and residual symmetries of the order parameter. New superconducting
phases and novel defects in the order parameter field, e.g. new types of vortices, are special
self-consistent solutions for the off-diagonal self-energies (gap functions) (see section II).
Collective modes in systems with unconventional pairing are discussed most thoroughly for triplet
pairing in $^3$He; we refer to the extensive literature in this field.\cite{vollhardt90,mck90} In
the following section we focus on collective effects induced by impurity scattering.

\section{Impurity Scattering}

\noindent Scattering by ordinary, nonmagnetic impurities is expected to have
many dramatic effects on the properties of unconventional
superconductors.\cite{gor87}
In this section we  discuss the theory of such effects,
using the  framework of the Fermi liquid theory of superconductivity.

In unconventional superconductors, it is useful to distinguish
two types of effects. Impurities affect global properties of
superconductors, such as thermodynamic properties ($T_{c}$,
critical fields, heat capacity,...) and long wavelength
response coefficients (conductivity, sound absorption,...).
Measurements of these macroscopic quantities average over the
positions of the impurities and give coarse-grained,
``impurity averaged'' information. Impurities also cause
strong local changes in their environment, out to distances
of order $\xi(T)$. Such local effects might be measured by
probes such as NMR and high resolution STM. These two different
effects require different types of calculations:
\begin{itemize}
\item Calculations which average over the positions of the
impurities, and so produce coarse-grained, ``impurity averaged''
quantities such as the transition temperature and density of
states.
\item Calculations which focus on the local environment of a single
impurity, and compute position-dependent quantities such
as the current or order parameter.
\end{itemize}
To start, we note that many of the properties of conventional
superconductors are very little affected by the presence of
nonmagnetic impurities. For example, quantities such as the
transition temperature, specific heat, and density of states
are quite insensitive to such scattering. On the other hand,
transport response functions such as the superfluid density
are affected rather strongly.

In contrast, in unconventional superconductors, all of the
quantities mentioned in the preceding paragraph are strongly
affected by nonmagnetic impurities. This opens up interesting
areas of research in two ways- novel effects in superconductivity
can be studied, and conversely, superconductivity can be used as a
probe of the properties of such impurities. We give some
illustrative examples:
\begin{enumerate}
\item Transition Temperature-- The value of $T_{c}$ is strongly
reduced by impurity scattering. The precise value of the
reduction depends on the angular dependence, in $\vec{k}$ space,
of both the order parameter and the scattering matrix.\cite{gor87}
If we 
let $\tau$ be the normal state transport time for  the
impurity scattering in question, then we can make a rough estimate
by saying that the reduction depends on the pair-breaking parameter
$1/T_{c}\tau$ in approximately the same way that the reduction
in conventional superconductors depends on $1/T_{c}\tau_{s}$,
where $\tau_{s}$ is the spin flip time.
\item Density of States-- $N(\omega)$ can be strongly affected.
For example, if the energy gap vanishes on a line of nodes
on the Fermi surface, for a pure superconductor we have
$N(\omega) \sim \omega$ for small energies. The addition of an
arbitrarily small concentration of impurities makes the
superconductor gapless;\cite{gor87,ued85}
this means that $N(\omega = 0)$ does not
vanish.
\item Order Parameter-- The question here is: how is the order
parameter $\Delta(\vec{k}_f,\vec{R})$ affected in the vicinity of a 
nonmagnetic impurity? In general $\Delta(\vec{k}_f,\vec{R})$
is  distorted
away from its pure state value $\Delta_{0}(\vec{k}_f)$.
However, in a conventional superconductor this distortion is
important only out to distances of order $1/k_{f}$ from the
impurity.\cite{fet65}
For an unconventional superconductor this distortion
has a much larger spatial extent, going out to distances of order
$\xi (T)$.\cite{rai77}
In addition, if the order parameter is a complex function
of $\vec{k}_f$, a pattern of supercurrents is set up near the impurity,
leading to the existence of a magnetic field;\cite{rai77,cho89a}
these currents will be
discussed later in this section.
Thus, what we are calling a ``nonmagnetic'' impurity can develop
magnetic properties in the superconducting state, if the
order parameter breaks time reversal symmetry. These properties
can be used as an experimental check of such an order parameter.
\item Superfluid Density Tensor-- In general, impurity scattering 
reduces the elements of $\bar{\bar{\rho}}_{s}$. One surprising effect
can occur in unconventional superconductors: impurity scattering can
increase the anisotropy in the $\rho_{s}$ tensor.\cite{cho88}
For example,
for an $l=1, m=0$
 order parameter with a line of nodes in the $xy$ plane, and a
spherical Fermi surface,
we can derive the following simple result close to $T_{c}$, as the
impurity scattering time $\tau$ goes to zero:
$\rho_{zz}/\rho_{xx} \rightarrow {-(10/3)} \ln(4\pi \tau T_{c})$.
\end{enumerate}

\subsection*{Impurity Averaging Calculations}

To illustrate the use of the impurity-averaged quasiclassical equation,
we review one particular calculation, a calculation which is 
perhaps somewhat novel. The general question to be addressed is:
if the magnitude of the energy gap varies through space, do
supercurrents arise? \cite{cho89,pal90a}
For weak-coupling theory in the absence of
impurities, the answer is no, if particle-hole symmetry 
is assumed.\cite{cro75}
That is, supercurrents are produced by spatial variations in the
phase and orientation of the order parameter, not by
variations in its magnitude.

 Note that if the gap is a complex
function of $\vec{k}$, and so breaks time reversal symmetry,
a term in the supercurrent proportional to the gradient of the
magnitude of the gap is allowed by symmetry. Gorkov\cite{gor87}
first pointed 
out that impurity scattering could give such a term a nonzero
coefficient, even in the context of a weak-coupling calculation.

For definiteness, we consider a singlet gap transforming
according to the $E_{1g}$ representation of the hexagonal group
$D_{6h}$. This particular case has been discussed as a possible model for the
heavy-fermion superconductor $UPt_{3}$.
\footnote{Similar calculations
to those in this section can be carried out for the E$_{2u}$
representation, which in our opinion is the most likely candidate for
the phases of UPt$_3$.}
Thus we write the gap as
$\Delta(\vec k_f)= \eta_{1}{\cal Y}_{1}(\vec k_f) + \eta_{2}{\cal Y}_{2}(\vec k_f)
$. 
Here, the ${\cal Y}_{i}(\vec k_f)$ are the basis functions, and the
coefficient $\vec{\eta} = (\eta_{1}, \eta_{2})$ transforms as
a vector in the $xy$ plane. In the
Ginzburg-Landau region, the supercurrent is given, in terms of
the gauge invariant derivative $\vec{D}=\vec{\nabla}+
\frac{2ie}{\hbar c}\vec{A}$, by\cite{cho89}
\begin{equation}
\hspace{-1.5em}\vec{J} = \frac{4e}{\hbar}Im[\kappa_{1}\eta_{j}\vec{D}^{*}\eta_{j}^{*}
+ \frac{1}{2}\kappa_{23}(\vec{\eta}D^{*}_{j}\eta^{*}_{j}
+\eta_{j}D^{*}_{j}\vec{\eta}^{*})
+\hat{z}\kappa_{4}\eta_{j}D^{*}_{z}\eta^{*}_{j}]
-\kappa_{a}\frac{ie}{\hbar}[\vec{\nabla}\times(\vec{\eta}\times\vec{\eta}^{*})]\,.
\end{equation}
This expression is obtained as 
the stationarity condition of the GL functional of eq. (\ref{free_energy}); the
coefficients $\kappa_{1}$, $\kappa_{23}$, $\kappa_{4}$, and $\kappa_{a}$ can
be calculated from the quasiclassical theory. 
We focus our attention on the coefficient $\kappa_{a}$; this is the
one which vanishes in weak-coupling theory unless impurity
scattering is included. Let us choose a particular form for the
order parameter:
$\vec{\eta}(\vec{R})=\eta_{0}(\vec{R})\vec{\phi}$,  
where $\vec{\phi}$ is a fixed two dimensional vector and $\eta_{0}
(\vec{R})$ is a spatially varying,
real amplitude. 
If we make the choice
$\vec{\phi} = (\hat{x} +i \hat{y})/\sqrt 2$
we then have
$\vec{J}= {(e\kappa_{a}}/{\hbar})\hat{z}\times\vec{\nabla}\eta_{0}^{2}$.

To calculate $\kappa_{a}$ we use the following strategy. We assume the
order parameter has the form given above, with an $\eta_{0}(\vec{R})$
which varies slowly in space. We then compute the Matsubara propagator\cite{ser83}
 $\hat{g}$ in a
gradient expansion($\hat{g}=\hat{g}^{(0)}+\hat{g}^{(1)}$),
use this $\hat{g}$ to compute the current,
and then read off $\kappa_{a}$ from our answer.
For simplicity we take our Fermi surface to be spherical ($\vec k_f=k_f \hat k$), and choose
the following basis functions:
${\cal Y}_{1}(\hat{k})=\surd 15 \hat{k}_{x} \hat{k}_{z}$, 
${\cal Y}_{2}(\hat{k})=\surd 15 \hat{k}_{y}\hat{k}_{z}$.
We evaluate the impurity self energy in the Born approximation, and
consider spherically symmetric impurities. This gives
\begin{equation}
\hat{\sigma}^{(j)}(\hat{k},\vec R;\epsilon_n)=
cN_f\int \frac{d^{2}\hat{k}^{\prime}}{4\pi}W(\hat{k},\hat{k}^{\prime})
\hat{g}^{(j)}(\hat{k}^{\prime},\vec R;\epsilon_n)\,.
\end{equation}
Here, $c$ is the concentration of impurities. We can expand $W$ in
spherical harmonics as follows:
\begin{equation}
W(\hat{k},\hat{k}^{\prime})=4\pi\sum_{l}W_{l}\sum_{m}
Y_{lm}^{*}(\hat{k})Y_{lm}(\hat{k}^{\prime})\,.
\end{equation}
Note that we are keeping all terms in the spherical harmonic
expansion of $W$; very often, calculations make the so-called
s-wave approximation, and keep only the $W_{0}$ term. As we shall
shortly see, the s-wave approximation is not good enough for our
purposes.

If we solve  for $\hat{g}^{(1)}$, and then compute
the current, we find the following formula 
for $\kappa_{a}$\cite{cho89,pal90a}
\begin{equation}
\kappa_{a} = \frac{N_f\xi^{2}_{0}}{80}\sum_{n\geq0}\frac{1}{n_+ 
(n_+ -\alpha_{2})^{2}}
\times\{\frac{\alpha_{1}}{(n_+-\alpha_{1})}-\frac
{\alpha_{3}}{(n_+ - \alpha_{3})}\} \,.
\end{equation}
Here, $n_+=n+1/2+\alpha_0$,   the pair breaking parameters $\alpha_{l}$ are defined by
$\alpha_{l}=cN_fW_{l}/2T_{c}$, 
and the correlation length is given by $\xi_{0}=
v_f/\pi T_{c}$.

There are several noteworthy features about this result for $\kappa_{a}$.
First, we see that it vanishes in the absence of impurity scattering.
In fact, if we had only kept the s-wave piece of $W$, we would still
have found that $\kappa_{a}$ was zero. Second, we can see that 
$\kappa_{a}$ depends on the details of the impurity scattering in a
quite intricate fashion. The $l=0,1,2,3$ pieces of $W$ all enter into the
answer. Finally, the sign of $\kappa_{a}$ can go either way. Note that
$\alpha_{0}$ must be positive, but the other $\alpha_{l}$'s can be
either positive or negative. 

We also remark that the coefficient $\kappa_{a}$ plays a role in the
discussion of whether or not Cooper pairs have angular momentum.
For a discussion of this point we refer to the 
literature.\cite{cho89,cro75,mer80}

\subsection*{Single Impurity Calculations}

Calculations concerning the local environment of a single impurity
are of interest for several reasons. On the one hand, experimentally
measurable quantities, such as local magnetic field, can be
investigated. These quantities can be important probes of the type
of order parameter present in the material. It is also true, however,
that these local types of calculation can be important in increasing
our physical insight;\cite{muz91} 
the impurity-averaging technique is
quite powerful, but the microscopic origin of interesting
effects may be hard to recover after the averaging procedure.

In an unconventional superconductor, much interesting structure
develops in the neighborhood of a nonmagnetic impurity. In contrast
to the conventional case, this structure is significant out to
quite long distances from the impurity, distances of order
$\xi(T)$. \cite{rai77}
If the order parameter breaks time reversal symmetry,
these local perturbations are particularly interesting:
a local pattern of equilibrium supercurrents is set up, leading to
the existence of a spatially varying magnetic field.\cite{cho89a}

We consider a single, nonmagnetic impurity, located at
$\vec{R}=0$; the interaction of the electronic quasiparticles
with the impurity can  be described by the potential
$v(\vec{k}_f,\vec{k}_f^{\prime})$. The quasiclassical approach \cite{thu81,thu84}
then leads to
coupled equations for the Matsubara propagator $\hat{g}(\vec{k}_f,\vec R;\epsilon_n)$,
the order parameter $\Delta(\vec{k}_f,\vec{R})$, and the t-matrix
$\hat{t}(\vec{k}_f,\vec{k}_f^{\prime};\epsilon_n)$. The equations for the propagator
are
\begin{equation}
[i\epsilon_n \hat{\tau}_{3}-\hat{\Delta},\hat{g}] +i\vec{v}_{f}
\cdot \vec{\nabla}\hat{g}=[\hat{t},\hat{g}_{imt}]\delta(\vec{R})
\  , \ \ \ \ 
\hat{g}\hat{g}=-\pi^{2}\hat 1\,.
\end{equation}
Here, the intermediate propagator $\hat{g}_{imt}$ satisfies the following 
equation:
$[i\epsilon_n\tau_{3}-\hat{\Delta},\hat{g}_{imt}]+i\vec{v}_{f}
\cdot \vec{\nabla}\hat{g}_{imt}=0
\ , \ \ \hat{g}_{imt}\hat{g}_{imt}=-\pi^{2}\hat 1$.
Note that the equations for both $\hat{g}$ and $\hat{g}_{imt}$
involve the order parameter $\Delta(\vec{k}_f,\vec{R})$, which must
be found from the gap equation; this gap equation needs $\hat{g}$ as input.

The final ingredient is an equation for the t-matrix:
\begin{equation}
\hspace{-1em}
\hat{t}(\vec{k}_f,\vec{k}_f^{\prime};\epsilon_n)= v(\vec{k}_f,\vec{k}_f^{\prime})\hat{1}+
\int \frac{d^{2}\vec{k}_f^{\prime \prime}}{(2\pi)^3\mid \vec v_f^{\prime \prime}\mid}\, 
v(\vec{k}_f,\vec{k}_f^{\prime \prime})\hat{g}_{imt}(\vec{k}_f^{\prime \prime},\vec R=0;\epsilon_n
)\hat{t}(\vec{k}_f^{\prime \prime},\vec{k}_f^{\prime};\epsilon_n)\,.
\end{equation}
So in general we face the difficult task of solving self-consistently
for all of the involved quantities. However, for a small
impurity, a certain simplification is possible.\cite{rai77,thu81} 
By small,
we mean an impurity with a cross section much smaller than
the square of the coherence length: $\sigma\ll \xi_{0}^{2}$. We may then
work to first order in the small parameter $\sigma/\xi^{2}_{0}$ and
write
\begin{equation}
\hat{g}= \hat{g}_{0}+\delta\hat{g}
\ , \ \ \ \hat{\Delta}=\hat{\Delta}_{0}+\delta\hat{\Delta}\,.
\end{equation}
Here, $\hat{g}_{0}$ and $\hat{\Delta}_{0}$ are the spatially
uniform, unperturbed propagator and order parameter, and we work to
first order in the small quantities $\delta\hat{g}$ and
$\delta\hat{\Delta}$. We may then take
$\hat{g}_{0}=\hat{g}_{imt}=
-\pi({i\epsilon_n \hat{\tau_{3}}-\hat{\Delta}_{0}})
/{(\epsilon_n^{2}+|\Delta_{0}|^{2})^{1/2}}
$.
The leading order answer for the Fourier
transform $\delta\hat{g}(\vec{k}_f,\vec q;\epsilon_n)$ is then\cite{cho90}
\begin{equation}
\delta\hat{g}(\vec{k}_f,\vec q;\epsilon_n)= \frac{E}{2\pi(E^{2}+Q^{2})}
(\hat{g}_{0}-\pi Q/E)([ \delta\hat{\Delta},\hat{g}_{0}]
+[\hat{t},\hat{g}_{0}])\,.
\end{equation}
Here, $Q=\frac{1}{2} \vec{v}_{f}\cdot \vec{q}$, and 
$E=(\epsilon_n^{2}+|\Delta_{0}|^{2})^{1/2}$.
This solution for $\delta \hat{g}$ can be used to calculate
quantities of direct physical interest, such as the supercurrent
or density of states. The off-diagonal, in particle-hole space,
elements of $\delta \hat{g}$ can be substituted into the gap
equation to obtain a closed equation for $\delta \Delta(\vec{k}_f,\vec{q})$.

We investigate in further detail the supercurrent. For any singlet,
or separable unitary triplet gap, we may write
$\delta\hat{\Delta}(\vec{k}_f,\vec{q})=
i\delta\Delta_{1}(\vec{k}_f,\vec{q})\hat{\tau}_{1} +i\delta\Delta_{2}(\vec{k}_f,\vec{q})
\hat{\tau}_{2}$.
We then obtain
\begin{equation}
\vec{J}(\vec{q})= -\frac{eT}{\pi}\sum_{\epsilon_n}\int
\frac{d^{2}\vec{k}_f}{(2\pi)^3\mid\vec v_f\mid}\, \vec{v}_{f}
\frac{E}{E^{2}+Q^{2}}I(\vec{k}_f,\vec{q};\epsilon_n)\,,
\end{equation}
\begin{equation}
I(\vec{k}_f,\vec{q};\epsilon_n)= \frac{2\pi^{2}iQ}{E^{2}}(\Delta_{2}\delta
\Delta_{1}-\Delta_{1}\delta\Delta_{2})+ \frac{1}{2}
Tr\left(\hat{\tau}_{3}\hat{g}_{0}[\hat{t},\hat{g}_{0}]\right)
-\frac{\pi Q}{2E}Tr\left(\hat{\tau}_{3}[\hat{t},\hat{g}_{0}]\right)\,.
\end{equation} 
We can now discuss the magnetic field, $B(\vec{R})$, and the total
magnetic moment $\vec{M}$, produced by these currents. We stress that
these currents vanish unless the unperturbed order parameter breaks
time reversal symmetry.  We ignore the Meissner screening current; this
is a good approximation for extreme type-II superconductors, since the
magnetic screening length is then  much longer than the spatial
extent of the currents.
\begin{itemize}
\item Magnetic Field $B$-- To estimate the overall size of the effect,
we consider the magnetic field produced at the impurity site,
$\vec{R}=0$. This is given by
\begin{equation}
\vec{B}(0)=\frac{1}{c}\int d^{3}R \frac{\vec{R}\times\vec{J}(\vec{R})}
{R^{3}} = \frac{4\pi i}{c}\int \frac{d^{3}q}{(2\pi)^{3}}
\frac{\vec{q}\times\vec{J}(\vec{q})}{q^{2}}\,.
\end{equation}

To make further progress in our estimates, we need the
expression for $\delta \Delta(\vec{k}_f,\vec{q})$. As mentioned above,
this is obtained by substituting $\delta\hat{g}$ into the gap
equation.
We refer to the literature for
these details,\cite{rai77,cho89a} 
and simply quote the final result:\cite{cho89a}
\begin{equation}
|\vec{B}(0)| \approx \beta(T)(ek_{f}^{2})(\frac{T_{c}}{T_{f}})^{2}
(\frac{v_f}{c})(\sigma k_{f}^{2})\,.
\end{equation}
In this formula, $\beta(T)$ is a temperature dependent function,
which vanishes at $T_{c}$, and is of order unity at lower
temperatures. The quantity $ek_{f}^{2}$ is a magnetic field, and
can be quite large, of order $10^{6} G$; however, it is reduced
by the small factors $(T_{c}/T_{f})^{2}$ and $(v_f/c)$.
\item Total Magnetic Moment $\vec{M}$-- The magnetic moment of the
currents around the impurity is given by
\begin{equation}
\vec{M}=\frac{1}{2c}\int d^{3}R\, \vec{R} \times \vec{J}(\vec{R})
=\frac{i}{2c}[\vec{\nabla}_{q} \times \vec{J}(\vec{q})]_{q=0}\,.
\end{equation}
The above formula for $\vec{J}(\vec{q})$ then allows us to
deduce a surprisingly general result:\cite{rai77,cho90} 
$\vec{M}=0$, unless 
$\delta\Delta(\vec{k}_f,\vec{q})$ is singular as $\vec{q}\rightarrow 0$.
This result holds for any $n(\vec{k}_f)$, $v(\vec{k}_f,\vec{k}_f^{\prime})$,$\Delta_{0}
(\vec{k}_f)$, and $\vec{v}_{f}$. 

The subtle point behind the $\vec{M}=0$ answer is the requirement
of nonsingular behavior at small $q$ of the perturbation in the
order parameter. This means that $\delta\Delta(\vec{k}_f,\vec{R})$
should go to zero quickly enough at large $R$. Now, the dangerous
distortions are the soft modes, those that leave the
bulk free energy unchanged. In a crystalline superconductor,
however, the orientation of the order parameter is
pinned by the crystal; thus, the only soft mode available is the
phase mode, which for real $\delta \theta(\vec{R})$ is
described by 
$\delta\Delta(\vec{k}_f,\vec{R}) = i\delta\theta(\vec{R})\Delta_{0}$.

So, we must study the gap equation for $\delta\Delta$ to see if the
impurity couples to the phase mode. No general proof exists,
but in all the cases studied so far, \cite{cho90}
a long range disturbance
in the phase has not been found.
\end{itemize} 
We hope that the preceding discussion gives a good idea of the
interesting spatial structure which should develop in the
neighborhood of an impurity in an unconventional superconductor.
It should be stressed that the length scale of this structure,
of order $\xi(T)$, is quite long by microscopic standards.

\section{Acknowledgements}

\noindent Part of this manuscript was written at the Institute for
Scientific Interchange in Torino, Italy. The authors DR and JAS thank
the ISI and the organizers for the workshop on `Phenomenological
Theories of Superconductivity' for their hospitality and support. The
work of JAS was supported in part by NSF grant no. DMR 91-20521 through
the Materials Research Center at Northwestern University, NSF
grant no. DMR 91-20000 through the Science and Technology Center
for Superconductivity, and the
Nordic Institute for Theoretical Physics in Copenhagen (NORDITA).


\begin{thebibliography}{86}%
\makeatletter
\providecommand \@ifxundefined [1]{%
 \@ifx{#1\undefined}
}%
\providecommand \@ifnum [1]{%
 \ifnum #1\expandafter \@firstoftwo
 \else \expandafter \@secondoftwo
 \fi
}%
\providecommand \@ifx [1]{%
 \ifx #1\expandafter \@firstoftwo
 \else \expandafter \@secondoftwo
 \fi
}%
\providecommand \natexlab [1]{#1}%
\providecommand \enquote  [1]{``#1''}%
\providecommand \bibnamefont  [1]{#1}%
\providecommand \bibfnamefont [1]{#1}%
\providecommand \citenamefont [1]{#1}%
\providecommand \href@noop [0]{\@secondoftwo}%
\providecommand \href [0]{\begingroup \@sanitize@url \@href}%
\providecommand \@href[1]{\@@startlink{#1}\@@href}%
\providecommand \@@href[1]{\endgroup#1\@@endlink}%
\providecommand \@sanitize@url [0]{\catcode `\\12\catcode `\$12\catcode
  `\&12\catcode `\#12\catcode `\^12\catcode `\_12\catcode `\%12\relax}%
\providecommand \@@startlink[1]{}%
\providecommand \@@endlink[0]{}%
\providecommand \url  [0]{\begingroup\@sanitize@url \@url }%
\providecommand \@url [1]{\endgroup\@href {#1}{\urlprefix }}%
\providecommand \urlprefix  [0]{URL }%
\providecommand \Eprint [0]{\href }%
\providecommand \doibase [0]{http://dx.doi.org/}%
\providecommand \selectlanguage [0]{\@gobble}%
\providecommand \bibinfo  [0]{\@secondoftwo}%
\providecommand \bibfield  [0]{\@secondoftwo}%
\providecommand \translation [1]{[#1]}%
\providecommand \BibitemOpen [0]{}%
\providecommand \bibitemStop [0]{}%
\providecommand \bibitemNoStop [0]{.\EOS\space}%
\providecommand \EOS [0]{\spacefactor3000\relax}%
\providecommand \BibitemShut  [1]{\csname bibitem#1\endcsname}%
\let\auto@bib@innerbib\@empty
\bibitem [{\citenamefont {Anderson}\ and\ \citenamefont {Morel}(1961)}]{and61}%
  \BibitemOpen
  \bibfield  {author} {\bibinfo {author} {\bibfnamefont {P.~W.}\ \bibnamefont
  {Anderson}}\ and\ \bibinfo {author} {\bibfnamefont {P.}~\bibnamefont
  {Morel}},\ }\bibfield  {{Generalized Bardeen-Cooper-Schrieffer States and the
  Proposed Low-Temperature Phase of $^3$He}} {\emph {\bibinfo {title}
  {{Generalized Bardeen-Cooper-Schrieffer States and the Proposed
  Low-Temperature Phase of $^3$He}},\ }}\href {\doibase
  10.1103/PhysRev.123.1911} {\bibfield  {journal} {\bibinfo  {journal} {Phys.
  Rev.}\ }\textbf {\bibinfo {volume} {123}},\ \bibinfo {pages} {1911} (\bibinfo
  {year} {1961})}\BibitemShut {NoStop}%
\bibitem [{\citenamefont {Osheroff}\ \emph {et~al.}(1972)\citenamefont
  {Osheroff}, \citenamefont {Richardson},\ and\ \citenamefont {Lee}}]{osh72}%
  \BibitemOpen
  \bibfield  {author} {\bibinfo {author} {\bibfnamefont {D.~D.}\ \bibnamefont
  {Osheroff}}, \bibinfo {author} {\bibfnamefont {R.~C.}\ \bibnamefont
  {Richardson}}, \ and\ \bibinfo {author} {\bibfnamefont {D.~M.}\ \bibnamefont
  {Lee}},\ }\bibfield  {{Evidence for a New Phase of Solid $^3$He}} {\emph
  {\bibinfo {title} {{Evidence for a New Phase of Solid $^3$He}},\ }}\href
  {\doibase 10.1103/PhysRevLett.28.885} {\bibfield  {journal} {\bibinfo
  {journal} {Phys. Rev. Lett.}\ }\textbf {\bibinfo {volume} {28}},\ \bibinfo
  {pages} {885} (\bibinfo {year} {1972})}\BibitemShut {NoStop}%
\bibitem [{\citenamefont {Steglich}\ \emph {et~al.}(1979)\citenamefont
  {Steglich}, \citenamefont {Aarts}, \citenamefont {Bredl}, \citenamefont
  {Lieke}, \citenamefont {Meschedes}, \citenamefont {Franz},\ and\
  \citenamefont {Sch\"afer}}]{ste79}%
  \BibitemOpen
  \bibfield  {author} {\bibinfo {author} {\bibfnamefont {F.}~\bibnamefont
  {Steglich}}, \bibinfo {author} {\bibfnamefont {J.}~\bibnamefont {Aarts}},
  \bibinfo {author} {\bibfnamefont {C.}~\bibnamefont {Bredl}}, \bibinfo
  {author} {\bibfnamefont {W.}~\bibnamefont {Lieke}}, \bibinfo {author}
  {\bibfnamefont {D.}~\bibnamefont {Meschedes}}, \bibinfo {author}
  {\bibfnamefont {W.}~\bibnamefont {Franz}}, \ and\ \bibinfo {author}
  {\bibfnamefont {H.}~\bibnamefont {Sch\"afer}},\ }\bibfield
  {{Superconductivity in the Presence of Strong Pauli Paramagnetism:
  CeCu$_2$Si$_2$ }} {\emph {\bibinfo {title} {{Superconductivity in the
  Presence of Strong Pauli Paramagnetism: CeCu$_2$Si$_2$ }},\ }}\href {\doibase
  10.1103/PhysRevLett.43.1892} {\bibfield  {journal} {\bibinfo  {journal}
  {Phys. Rev. Lett.}\ }\textbf {\bibinfo {volume} {43}},\ \bibinfo {pages}
  {1892} (\bibinfo {year} {1979})}\BibitemShut {NoStop}%
\bibitem [{\citenamefont {Ott}\ \emph {et~al.}(1987)\citenamefont {Ott},
  \citenamefont {Felder}, \citenamefont {Bernasconi}, \citenamefont {Fisk},
  \citenamefont {Smith}, \citenamefont {Taillefer},\ and\ \citenamefont
  {Lonzarich}}]{ott87}%
  \BibitemOpen
  \bibfield  {author} {\bibinfo {author} {\bibfnamefont {H.}~\bibnamefont
  {Ott}}, \bibinfo {author} {\bibfnamefont {E.}~\bibnamefont {Felder}},
  \bibinfo {author} {\bibfnamefont {A.}~\bibnamefont {Bernasconi}}, \bibinfo
  {author} {\bibfnamefont {Z.}~\bibnamefont {Fisk}}, \bibinfo {author}
  {\bibfnamefont {J.}~\bibnamefont {Smith}}, \bibinfo {author} {\bibfnamefont
  {L.}~\bibnamefont {Taillefer}}, \ and\ \bibinfo {author} {\bibfnamefont
  {G.}~\bibnamefont {Lonzarich}},\ }\bibfield  {{Specific Heat and Thermal
  Conductivity of Superconducting $UBe_{13}$ and $UPt_{3}$ at Very Low
  Temperatures}} {\emph {\bibinfo {title} {{Specific Heat and Thermal
  Conductivity of Superconducting $UBe_{13}$ and $UPt_{3}$ at Very Low
  Temperatures}},\ }}\href {\doibase 10.7567/JJAPS.26S3.1217} {\bibfield
  {journal} {\bibinfo  {journal} {Japanese Journal of Applied Physics}\
  }\textbf {\bibinfo {volume} {26}},\ \bibinfo {pages} {1217} (\bibinfo {year}
  {1987})},\ \bibinfo {note} {18th International Conference on Low Temperature
  Physics, Kyoto}\BibitemShut {NoStop}%
\bibitem [{\citenamefont {Sarma}\ \emph {et~al.}(1992)\citenamefont {Sarma},
  \citenamefont {Levy}, \citenamefont {Adenwalla},\ and\ \citenamefont
  {Ketterson}}]{sar92}%
  \BibitemOpen
  \bibfield  {author} {\bibinfo {author} {\bibfnamefont {B.}~\bibnamefont
  {Sarma}}, \bibinfo {author} {\bibfnamefont {M.}~\bibnamefont {Levy}},
  \bibinfo {author} {\bibfnamefont {S.}~\bibnamefont {Adenwalla}}, \ and\
  \bibinfo {author} {\bibfnamefont {J.}~\bibnamefont {Ketterson}},\ }\bibinfo
  {title} {Sound propagation in the heavy fermion superconductors},\ in\
  \href@noop {} {\emph {\bibinfo {booktitle} {Physical Acoustics}}},\
  Vol.~\bibinfo {volume} {XX}\ (\bibinfo  {publisher} {Academic Press},\
  \bibinfo {address} {New York},\ \bibinfo {year} {1992})\ p.\ \bibinfo {pages}
  {107}\BibitemShut {NoStop}%
\bibitem [{\citenamefont {Steglich}\ and\ \citenamefont {{\it
  et~al.}}(1993)}]{ste93}%
  \BibitemOpen
  \bibfield  {author} {\bibinfo {author} {\bibfnamefont {F.}~\bibnamefont
  {Steglich}}\ and\ \bibinfo {author} {\bibnamefont {{\it et~al.}}},\
  }\href@noop {} {\bibfield  {journal} {\bibinfo  {journal} {Frontiers Sol.
  State Sciences}\ }\textbf {\bibinfo {volume} {1}},\ \bibinfo {pages} {527}
  (\bibinfo {year} {1993})}\BibitemShut {NoStop}%
\bibitem [{\citenamefont {Bickers}\ \emph {et~al.}(1987)\citenamefont
  {Bickers}, \citenamefont {Scalapino},\ and\ \citenamefont
  {Scalettar}}]{bic87}%
  \BibitemOpen
  \bibfield  {author} {\bibinfo {author} {\bibfnamefont {N.~E.}\ \bibnamefont
  {Bickers}}, \bibinfo {author} {\bibfnamefont {D.~J.}\ \bibnamefont
  {Scalapino}}, \ and\ \bibinfo {author} {\bibfnamefont {R.~T.}\ \bibnamefont
  {Scalettar}},\ }\bibfield  {{CDW and SDW mediated pairing interactions}}
  {\emph {\bibinfo {title} {{CDW and SDW mediated pairing interactions}},\
  }}\href {\doibase 10.1142/S0217979287001079} {\bibfield  {journal} {\bibinfo
  {journal} {Int. Journ. Mod. Phys. B}\ }\textbf {\bibinfo {volume} {1}},\
  \bibinfo {pages} {687} (\bibinfo {year} {1987})}\BibitemShut {NoStop}%
\bibitem [{\citenamefont {Lee}\ \emph {et~al.}(1986)\citenamefont {Lee},
  \citenamefont {Rice}, \citenamefont {Serene}, \citenamefont {Sham},\ and\
  \citenamefont {Wilkins}}]{lee86}%
  \BibitemOpen
  \bibfield  {author} {\bibinfo {author} {\bibfnamefont {P.~A.}\ \bibnamefont
  {Lee}}, \bibinfo {author} {\bibfnamefont {T.~M.}\ \bibnamefont {Rice}},
  \bibinfo {author} {\bibfnamefont {J.~W.}\ \bibnamefont {Serene}}, \bibinfo
  {author} {\bibfnamefont {L.~J.}\ \bibnamefont {Sham}}, \ and\ \bibinfo
  {author} {\bibfnamefont {J.~W.}\ \bibnamefont {Wilkins}},\ }\bibfield
  {{Theories of Heavy-Electron Systems}} {\emph {\bibinfo {title} {{Theories of
  Heavy-Electron Systems}},\ }}\href@noop {} {\bibfield  {journal} {\bibinfo
  {journal} {Comments on Cond. Mat. Phys.}\ }\textbf {\bibinfo {volume} {12}},\
  \bibinfo {pages} {99} (\bibinfo {year} {1986})}\BibitemShut {NoStop}%
\bibitem [{\citenamefont {Gor'kov}(1987)}]{gor87}%
  \BibitemOpen
  \bibfield  {author} {\bibinfo {author} {\bibfnamefont {L.}~\bibnamefont
  {Gor'kov}},\ }\bibfield  {{Superconductivity in heavy fermion systems}}
  {\emph {\bibinfo {title} {{Superconductivity in heavy fermion systems}},\
  }}\href@noop {} {\bibfield  {journal} {\bibinfo  {journal} {Sov. Sci. Rev.
  A.}\ }\textbf {\bibinfo {volume} {9}},\ \bibinfo {pages} {1} (\bibinfo {year}
  {1987})}\BibitemShut {NoStop}%
\bibitem [{\citenamefont {Sigrist}\ and\ \citenamefont {Ueda}(1991)}]{sig91}%
  \BibitemOpen
  \bibfield  {author} {\bibinfo {author} {\bibfnamefont {M.}~\bibnamefont
  {Sigrist}}\ and\ \bibinfo {author} {\bibfnamefont {K.}~\bibnamefont {Ueda}},\
  }\bibfield  {{Phenomenological Theories of Unconventional Superconductivity}}
  {\emph {\bibinfo {title} {{Phenomenological Theories of Unconventional
  Superconductivity}},\ }}\href {\doibase 10.1103/RevModPhys.63.239} {\bibfield
   {journal} {\bibinfo  {journal} {Rev. Mod. Phys.}\ }\textbf {\bibinfo
  {volume} {63}},\ \bibinfo {pages} {239} (\bibinfo {year} {1991})}\BibitemShut
  {NoStop}%
\bibitem [{Note1()}]{Note1}%
  \BibitemOpen
  \bibinfo {note} {In the heavy-fermion materials it is generally assumed that
  spin-orbit coupling is strong;\cite {and84a,vol85,lee86} thus, the labels
  characterizing the quasiparticle states are not eigenvalues of the spin
  operator for electrons. Nevertheless, in zero-field the Kramers degeneracy
  guarantees that each $\protect \vec {k}$ state is two-fold degenerate, and
  thus, may be labeled by a `pseudo-spin' quantum number $\alpha $, which can
  take on two possible values. We use the term `spin' interchangeably with
  `pseudo-spin' in this article.}\BibitemShut {Stop}%
\bibitem [{\citenamefont {Anderson}(1984)}]{and84a}%
  \BibitemOpen
  \bibfield  {author} {\bibinfo {author} {\bibfnamefont {P.~W.}\ \bibnamefont
  {Anderson}},\ }\bibfield  {{Structure of "triplet" superconducting energy
  gaps}} {\emph {\bibinfo {title} {{Structure of "triplet" superconducting
  energy gaps}},\ }}\href {\doibase 10.1103/PhysRevB.30.4000} {\bibfield
  {journal} {\bibinfo  {journal} {Phys. Rev. B}\ }\textbf {\bibinfo {volume}
  {30}},\ \bibinfo {pages} {4000} (\bibinfo {year} {1984})}\BibitemShut
  {NoStop}%
\bibitem [{\citenamefont {Volovik}\ and\ \citenamefont
  {Gor'kov}(1985)}]{vol85}%
  \BibitemOpen
  \bibfield  {author} {\bibinfo {author} {\bibfnamefont {G.}~\bibnamefont
  {Volovik}}\ and\ \bibinfo {author} {\bibfnamefont {L.}~\bibnamefont
  {Gor'kov}},\ }\bibfield  {{Superconducting Classes in Heavy Fermion Systems}}
  {\emph {\bibinfo {title} {{Superconducting Classes in Heavy Fermion
  Systems}},\ }}\href {\doibase 10.1007/978-94-011-1622-0_14} {\bibfield
  {journal} {\bibinfo  {journal} {Sov. Phys. JETP}\ }\textbf {\bibinfo {volume}
  {61}},\ \bibinfo {pages} {843} (\bibinfo {year} {1985})}\BibitemShut
  {NoStop}%
\bibitem [{\citenamefont {Yip}\ and\ \citenamefont {Garg}(1993)}]{yip93c}%
  \BibitemOpen
  \bibfield  {author} {\bibinfo {author} {\bibfnamefont {S.}~\bibnamefont
  {Yip}}\ and\ \bibinfo {author} {\bibfnamefont {A.}~\bibnamefont {Garg}},\
  }\bibfield  {Superconducting states of reduced symmetry: General order
  parameters and physical implications} {\emph {\bibinfo {title}
  {Superconducting states of reduced symmetry: General order parameters and
  physical implications},\ }}\href {\doibase 10.1103/PhysRevB.48.3304}
  {\bibfield  {journal} {\bibinfo  {journal} {Phys. Rev. B}\ }\textbf {\bibinfo
  {volume} {48}},\ \bibinfo {pages} {3304} (\bibinfo {year}
  {1993})}\BibitemShut {NoStop}%
\bibitem [{\citenamefont {M\"uller}\ \emph {et~al.}(1987)\citenamefont
  {M\"uller}, \citenamefont {Roth}, \citenamefont {Maurer}, \citenamefont
  {Scheidt}, \citenamefont {L\"uders}, \citenamefont {Bucher},\ and\
  \citenamefont {B\"ommel}}]{mul87}%
  \BibitemOpen
  \bibfield  {author} {\bibinfo {author} {\bibfnamefont {V.}~\bibnamefont
  {M\"uller}}, \bibinfo {author} {\bibfnamefont {C.}~\bibnamefont {Roth}},
  \bibinfo {author} {\bibfnamefont {D.}~\bibnamefont {Maurer}}, \bibinfo
  {author} {\bibfnamefont {E.}~\bibnamefont {Scheidt}}, \bibinfo {author}
  {\bibfnamefont {K.}~\bibnamefont {L\"uders}}, \bibinfo {author}
  {\bibfnamefont {E.}~\bibnamefont {Bucher}}, \ and\ \bibinfo {author}
  {\bibfnamefont {H.}~\bibnamefont {B\"ommel}},\ }\bibfield  {{Ultrasonic
  determination of different phases in superconducting UPt$_3$}} {\emph
  {\bibinfo {title} {{Ultrasonic determination of different phases in
  superconducting UPt$_3$}},\ }}\href {\doibase 10.1103/PhysRevLett.58.1224}
  {\bibfield  {journal} {\bibinfo  {journal} {Phys. Rev. Lett.}\ }\textbf
  {\bibinfo {volume} {58}},\ \bibinfo {pages} {1224} (\bibinfo {year}
  {1987})}\BibitemShut {NoStop}%
\bibitem [{\citenamefont {Qian}\ \emph {et~al.}(1987)\citenamefont {Qian},
  \citenamefont {Xu}, \citenamefont {Schenstrom}, \citenamefont {Baum},
  \citenamefont {Ketterson}, \citenamefont {Hinks}, \citenamefont {Levy},\ and\
  \citenamefont {Sarma}}]{qia87}%
  \BibitemOpen
  \bibfield  {author} {\bibinfo {author} {\bibfnamefont {Y.}~\bibnamefont
  {Qian}}, \bibinfo {author} {\bibfnamefont {M.-F.}\ \bibnamefont {Xu}},
  \bibinfo {author} {\bibfnamefont {A.}~\bibnamefont {Schenstrom}}, \bibinfo
  {author} {\bibfnamefont {H.-P.}\ \bibnamefont {Baum}}, \bibinfo {author}
  {\bibfnamefont {J.}~\bibnamefont {Ketterson}}, \bibinfo {author}
  {\bibfnamefont {D.}~\bibnamefont {Hinks}}, \bibinfo {author} {\bibfnamefont
  {M.}~\bibnamefont {Levy}}, \ and\ \bibinfo {author} {\bibfnamefont
  {B.}~\bibnamefont {Sarma}},\ }\bibfield  {{Longitudinal sound measurements on
  UPt$_3$ in a magnetic field}} {\emph {\bibinfo {title} {{Longitudinal sound
  measurements on UPt$_3$ in a magnetic field}},\ }}\href {\doibase
  10.1016/0038-1098(87)90861-1} {\bibfield  {journal} {\bibinfo  {journal}
  {Solid State Commun.}\ }\textbf {\bibinfo {volume} {63}},\ \bibinfo {pages}
  {599} (\bibinfo {year} {1987})}\BibitemShut {NoStop}%
\bibitem [{\citenamefont {Schenstrom}\ \emph {et~al.}(1989)\citenamefont
  {Schenstrom}, \citenamefont {Xu}, \citenamefont {Hong}, \citenamefont {Bein},
  \citenamefont {Levy}, \citenamefont {Sarma}, \citenamefont {Adenwalla},
  \citenamefont {Zhao}, \citenamefont {Tokuyasu}, \citenamefont {Hess},
  \citenamefont {Ketterson}, \citenamefont {Sauls},\ and\ \citenamefont
  {Hinks}}]{sch89}%
  \BibitemOpen
  \bibfield  {author} {\bibinfo {author} {\bibfnamefont {A.}~\bibnamefont
  {Schenstrom}}, \bibinfo {author} {\bibfnamefont {M.-F.}\ \bibnamefont {Xu}},
  \bibinfo {author} {\bibfnamefont {Y.}~\bibnamefont {Hong}}, \bibinfo {author}
  {\bibfnamefont {D.}~\bibnamefont {Bein}}, \bibinfo {author} {\bibfnamefont
  {M.}~\bibnamefont {Levy}}, \bibinfo {author} {\bibfnamefont {B.}~\bibnamefont
  {Sarma}}, \bibinfo {author} {\bibfnamefont {S.}~\bibnamefont {Adenwalla}},
  \bibinfo {author} {\bibfnamefont {Z.}~\bibnamefont {Zhao}}, \bibinfo {author}
  {\bibfnamefont {T.}~\bibnamefont {Tokuyasu}}, \bibinfo {author}
  {\bibfnamefont {D.~W.}\ \bibnamefont {Hess}}, \bibinfo {author}
  {\bibfnamefont {J.~B.}\ \bibnamefont {Ketterson}}, \bibinfo {author}
  {\bibfnamefont {J.~A.}\ \bibnamefont {Sauls}}, \ and\ \bibinfo {author}
  {\bibfnamefont {D.~G.}\ \bibnamefont {Hinks}},\ }\bibfield  {{Anisotropy of
  the Magnetic-Field-Induced Phase Transition in Superconducting UPt$_3$}}
  {\emph {\bibinfo {title} {{Anisotropy of the Magnetic-Field-Induced Phase
  Transition in Superconducting UPt$_3$}},\ }}\href {\doibase
  10.1103/PhysRevLett.62.332} {\bibfield  {journal} {\bibinfo  {journal} {Phys.
  Rev. Lett.}\ }\textbf {\bibinfo {volume} {62}},\ \bibinfo {pages} {332}
  (\bibinfo {year} {1989})}\BibitemShut {NoStop}%
\bibitem [{\citenamefont {Bruls}\ \emph {et~al.}(1990)\citenamefont {Bruls},
  \citenamefont {Weber}, \citenamefont {Wolf}, \citenamefont {Thalmeier},
  \citenamefont {Luthi}, \citenamefont {de~Visser},\ and\ \citenamefont
  {Menovsky}}]{bru90}%
  \BibitemOpen
  \bibfield  {author} {\bibinfo {author} {\bibfnamefont {G.}~\bibnamefont
  {Bruls}}, \bibinfo {author} {\bibfnamefont {D.}~\bibnamefont {Weber}},
  \bibinfo {author} {\bibfnamefont {B.}~\bibnamefont {Wolf}}, \bibinfo {author}
  {\bibfnamefont {P.}~\bibnamefont {Thalmeier}}, \bibinfo {author}
  {\bibfnamefont {B.}~\bibnamefont {Luthi}}, \bibinfo {author} {\bibfnamefont
  {A.}~\bibnamefont {de~Visser}}, \ and\ \bibinfo {author} {\bibfnamefont
  {A.}~\bibnamefont {Menovsky}},\ }\bibfield  {{Strain–order-parameter
  coupling and phase diagrams in superconducting UPt$_3$}} {\emph {\bibinfo
  {title} {{Strain–order-parameter coupling and phase diagrams in
  superconducting UPt$_3$}},\ }}\href {\doibase 10.1103/PhysRevLett.65.2294}
  {\bibfield  {journal} {\bibinfo  {journal} {Phys. Rev. Lett.}\ }\textbf
  {\bibinfo {volume} {65}},\ \bibinfo {pages} {2294} (\bibinfo {year}
  {1990})}\BibitemShut {NoStop}%
\bibitem [{\citenamefont {Adenwalla}\ \emph {et~al.}(1990)\citenamefont
  {Adenwalla}, \citenamefont {Lin}, \citenamefont {Ran}, \citenamefont {Zhao},
  \citenamefont {Ketterson}, \citenamefont {Sauls}, \citenamefont {Taillefer},
  \citenamefont {Hinks}, \citenamefont {Levy},\ and\ \citenamefont
  {Sarma}}]{ade90}%
  \BibitemOpen
  \bibfield  {author} {\bibinfo {author} {\bibfnamefont {S.}~\bibnamefont
  {Adenwalla}}, \bibinfo {author} {\bibfnamefont {S.}~\bibnamefont {Lin}},
  \bibinfo {author} {\bibfnamefont {Q.}~\bibnamefont {Ran}}, \bibinfo {author}
  {\bibfnamefont {Z.}~\bibnamefont {Zhao}}, \bibinfo {author} {\bibfnamefont
  {J.}~\bibnamefont {Ketterson}}, \bibinfo {author} {\bibfnamefont
  {J.}~\bibnamefont {Sauls}}, \bibinfo {author} {\bibfnamefont
  {L.}~\bibnamefont {Taillefer}}, \bibinfo {author} {\bibfnamefont
  {D.}~\bibnamefont {Hinks}}, \bibinfo {author} {\bibfnamefont
  {M.}~\bibnamefont {Levy}}, \ and\ \bibinfo {author} {\bibfnamefont
  {B.}~\bibnamefont {Sarma}},\ }\bibfield  {{Phase diagram of UPt$_3$ from
  ultrasonic velocity measurements}} {\emph {\bibinfo {title} {{Phase diagram
  of UPt$_3$ from ultrasonic velocity measurements}},\ }}\href {\doibase
  10.1103/PhysRevLett.65.2298} {\bibfield  {journal} {\bibinfo  {journal}
  {Phys. Rev. Lett.}\ }\textbf {\bibinfo {volume} {65}},\ \bibinfo {pages}
  {2298} (\bibinfo {year} {1990})}\BibitemShut {NoStop}%
\bibitem [{\citenamefont {Hess}\ \emph {et~al.}(1989)\citenamefont {Hess},
  \citenamefont {Tokuyasu},\ and\ \citenamefont {Sauls}}]{hes89}%
  \BibitemOpen
  \bibfield  {author} {\bibinfo {author} {\bibfnamefont {D.}~\bibnamefont
  {Hess}}, \bibinfo {author} {\bibfnamefont {T.~A.}\ \bibnamefont {Tokuyasu}},
  \ and\ \bibinfo {author} {\bibfnamefont {J.~A.}\ \bibnamefont {Sauls}},\
  }\bibfield  {{Broken Symmetry in an Unconventional Superconductor: A Model
  for the Double Transition in UPt$_3$}} {\emph {\bibinfo {title} {{Broken
  Symmetry in an Unconventional Superconductor: A Model for the Double
  Transition in UPt$_3$}},\ }}\href {\doibase 10.1088/0953-8984/1/43/014}
  {\bibfield  {journal} {\bibinfo  {journal} {J. Phys. Cond. Matt.}\ }\textbf
  {\bibinfo {volume} {1}},\ \bibinfo {pages} {8135} (\bibinfo {year}
  {1989})}\BibitemShut {NoStop}%
\bibitem [{\citenamefont {Machida}\ and\ \citenamefont {Ozaki}(1989)}]{mac89}%
  \BibitemOpen
  \bibfield  {author} {\bibinfo {author} {\bibfnamefont {K.}~\bibnamefont
  {Machida}}\ and\ \bibinfo {author} {\bibfnamefont {M.}~\bibnamefont
  {Ozaki}},\ }\bibfield  {{Splitting of Superconducting Transitions in
  UPt$_3$}} {\emph {\bibinfo {title} {{Splitting of Superconducting Transitions
  in UPt$_3$}},\ }}\href {\doibase 10.1143/JPSJ.58.2244} {\bibfield  {journal}
  {\bibinfo  {journal} {J. Phys. Soc. Jpn}\ }\textbf {\bibinfo {volume} {58}},\
  \bibinfo {pages} {2244} (\bibinfo {year} {1989})}\BibitemShut {NoStop}%
\bibitem [{\citenamefont {Sauls}(1994{\natexlab{a}})}]{sau94b}%
  \BibitemOpen
  \bibfield  {author} {\bibinfo {author} {\bibfnamefont {J.~A.}\ \bibnamefont
  {Sauls}},\ }\bibfield  {{A Theory for the Superconducting Phases of UPt$_3$}}
  {\emph {\bibinfo {title} {{A Theory for the Superconducting Phases of
  UPt$_3$}},\ }}\href {\doibase 10.1007/BF00754932} {\bibfield  {journal}
  {\bibinfo  {journal} {J. Low Temp. Phys.}\ }\textbf {\bibinfo {volume}
  {95}},\ \bibinfo {pages} {153} (\bibinfo {year} {1994}{\natexlab{a}})},\
  \bibinfo {note} {{Cologne Workshop in Honor of Dieter Wohleben -
  1993}}\BibitemShut {NoStop}%
\bibitem [{\citenamefont {Sauls}(1994{\natexlab{b}})}]{sau94}%
  \BibitemOpen
  \bibfield  {author} {\bibinfo {author} {\bibfnamefont {J.~A.}\ \bibnamefont
  {Sauls}},\ }\bibfield  {The Order Parameter for the Superconducting Phases of
  {UPt$_3$}} {\emph {\bibinfo {title} {The order parameter for the
  superconducting phases of {UPt$_3$}},\ }}\href
  {https://arxiv.org/abs/1812.09984} {\bibfield  {journal} {\bibinfo  {journal}
  {Adv. Phys.}\ }\textbf {\bibinfo {volume} {43}},\ \bibinfo {pages} {113}
  (\bibinfo {year} {1994}{\natexlab{b}})}\BibitemShut {NoStop}%
\bibitem [{\citenamefont {Joynt}(1988)}]{joy88}%
  \BibitemOpen
  \bibfield  {author} {\bibinfo {author} {\bibfnamefont {R.}~\bibnamefont
  {Joynt}},\ }\bibfield  {{Phase diagram of d-wave superconductors in a
  magnetic field}} {\emph {\bibinfo {title} {{Phase diagram of d-wave
  superconductors in a magnetic field}},\ }}\href {\doibase
  10.1088/0953-2048/1/4/012} {\bibfield  {journal} {\bibinfo  {journal} {Sup.
  Sci. Tech.}\ }\textbf {\bibinfo {volume} {1}},\ \bibinfo {pages} {210}
  (\bibinfo {year} {1988})}\BibitemShut {NoStop}%
\bibitem [{\citenamefont {Aeppli}\ \emph {et~al.}(1988)\citenamefont {Aeppli},
  \citenamefont {Bucher}, \citenamefont {Broholm}, \citenamefont {Kjems},
  \citenamefont {Baumann},\ and\ \citenamefont {Hufnagl}}]{aep88}%
  \BibitemOpen
  \bibfield  {author} {\bibinfo {author} {\bibfnamefont {G.}~\bibnamefont
  {Aeppli}}, \bibinfo {author} {\bibfnamefont {E.}~\bibnamefont {Bucher}},
  \bibinfo {author} {\bibfnamefont {C.}~\bibnamefont {Broholm}}, \bibinfo
  {author} {\bibfnamefont {J.}~\bibnamefont {Kjems}}, \bibinfo {author}
  {\bibfnamefont {J.}~\bibnamefont {Baumann}}, \ and\ \bibinfo {author}
  {\bibfnamefont {J.}~\bibnamefont {Hufnagl}},\ }\bibfield  {{Magnetic order
  and fluctuations in superconducting UPt$_3$}} {\emph {\bibinfo {title}
  {{Magnetic order and fluctuations in superconducting UPt$_3$}},\ }}\href
  {\doibase 10.1103/PhysRevLett.60.615} {\bibfield  {journal} {\bibinfo
  {journal} {Phys. Rev. Lett.}\ }\textbf {\bibinfo {volume} {60}},\ \bibinfo
  {pages} {615} (\bibinfo {year} {1988})}\BibitemShut {NoStop}%
\bibitem [{\citenamefont {Vorenkamp}(1992)}]{vor92}%
  \BibitemOpen
  \bibfield  {author} {\bibinfo {author} {\bibfnamefont {T.}~\bibnamefont
  {Vorenkamp}},\ }\emph {\bibinfo {title} {"Unconventional Superconductivity in
  Heavy Fermion Compound UPt$_3$}},\ \href@noop {} {Ph.D. thesis},\ \bibinfo
  {school} {University of Amsterdam} (\bibinfo {year} {1992})\BibitemShut
  {NoStop}%
\bibitem [{\citenamefont {Shivaram}\ \emph {et~al.}(1989)\citenamefont
  {Shivaram}, \citenamefont {{Gannon Jr.}},\ and\ \citenamefont
  {Hinks}}]{shi89}%
  \BibitemOpen
  \bibfield  {author} {\bibinfo {author} {\bibfnamefont {B.}~\bibnamefont
  {Shivaram}}, \bibinfo {author} {\bibfnamefont {J.}~\bibnamefont {{Gannon
  Jr.}}}, \ and\ \bibinfo {author} {\bibfnamefont {D.}~\bibnamefont {Hinks}},\
  }\bibfield  {{Lower and upper critical fields in the heavy electron
  superconductor UPt$_3$}} {\emph {\bibinfo {title} {{Lower and upper critical
  fields in the heavy electron superconductor UPt$_3$}},\ }}\href {\doibase
  10.1103/PhysRevLett.63.1723} {\bibfield  {journal} {\bibinfo  {journal}
  {Phys. Rev. Lett.}\ }\textbf {\bibinfo {volume} {63}},\ \bibinfo {pages}
  {1723} (\bibinfo {year} {1989})}\BibitemShut {NoStop}%
\bibitem [{\citenamefont {Vincent}\ \emph {et~al.}(1991)\citenamefont
  {Vincent}, \citenamefont {Hammann}, \citenamefont {Taillefer}, \citenamefont
  {Behnia}, \citenamefont {Keller},\ and\ \citenamefont {Flouquet}}]{vin91}%
  \BibitemOpen
  \bibfield  {author} {\bibinfo {author} {\bibfnamefont {E.}~\bibnamefont
  {Vincent}}, \bibinfo {author} {\bibfnamefont {J.}~\bibnamefont {Hammann}},
  \bibinfo {author} {\bibfnamefont {L.}~\bibnamefont {Taillefer}}, \bibinfo
  {author} {\bibfnamefont {K.}~\bibnamefont {Behnia}}, \bibinfo {author}
  {\bibfnamefont {N.}~\bibnamefont {Keller}}, \ and\ \bibinfo {author}
  {\bibfnamefont {J.}~\bibnamefont {Flouquet}},\ }\bibfield  {{Low-field
  diamagnetic response of the superconducting phases in UPt$_3$}} {\emph
  {\bibinfo {title} {{Low-field diamagnetic response of the superconducting
  phases in UPt$_3$}},\ }}\href {\doibase 10.1088/0953-8984/3/20/013}
  {\bibfield  {journal} {\bibinfo  {journal} {J. Phys. Cond. Matt.}\ }\textbf
  {\bibinfo {volume} {3}},\ \bibinfo {pages} {3517} (\bibinfo {year}
  {1991})}\BibitemShut {NoStop}%
\bibitem [{\citenamefont {Knetch}\ \emph {et~al.}(1992)\citenamefont {Knetch},
  \citenamefont {Mydosh}, \citenamefont {Vorenkamp},\ and\ \citenamefont
  {Menovsky}}]{kne92}%
  \BibitemOpen
  \bibfield  {author} {\bibinfo {author} {\bibfnamefont {E.~A.}\ \bibnamefont
  {Knetch}}, \bibinfo {author} {\bibfnamefont {J.~A.}\ \bibnamefont {Mydosh}},
  \bibinfo {author} {\bibfnamefont {T.}~\bibnamefont {Vorenkamp}}, \ and\
  \bibinfo {author} {\bibfnamefont {A.~A.}\ \bibnamefont {Menovsky}},\
  }\bibfield  {{Temperature dependence of the lower critical field of UPt$_3$
  doped with boron}} {\emph {\bibinfo {title} {{Temperature dependence of the
  lower critical field of UPt$_3$ doped with boron}},\ }}\href {\doibase
  10.1016/0304-8853(92)91355-W} {\bibfield  {journal} {\bibinfo  {journal} {J.
  Mag. Mag. Mat.}\ }\textbf {\bibinfo {volume} {108}},\ \bibinfo {pages} {75}
  (\bibinfo {year} {1992})}\BibitemShut {NoStop}%
\bibitem [{\citenamefont {Trappmann}\ \emph {et~al.}(1991)\citenamefont
  {Trappmann}, \citenamefont {v.~L\"ohneysen},\ and\ \citenamefont
  {Taillefer}}]{tra91}%
  \BibitemOpen
  \bibfield  {author} {\bibinfo {author} {\bibfnamefont {T.}~\bibnamefont
  {Trappmann}}, \bibinfo {author} {\bibfnamefont {H.}~\bibnamefont
  {v.~L\"ohneysen}}, \ and\ \bibinfo {author} {\bibfnamefont {L.}~\bibnamefont
  {Taillefer}},\ }\bibfield  {{Pressure dependence of the superconducting
  phases in UPt$_3$}} {\emph {\bibinfo {title} {{Pressure dependence of the
  superconducting phases in UPt$_3$}},\ }}\href {\doibase
  10.1103/PhysRevB.43.13714} {\bibfield  {journal} {\bibinfo  {journal} {Phys.
  Rev. B}\ }\textbf {\bibinfo {volume} {43}},\ \bibinfo {pages} {13714}
  (\bibinfo {year} {1991})}\BibitemShut {NoStop}%
\bibitem [{\citenamefont {Hayden}\ \emph {et~al.}(1992)\citenamefont {Hayden},
  \citenamefont {Taillefer}, \citenamefont {Vettier},\ and\ \citenamefont
  {Flouquet}}]{hay92}%
  \BibitemOpen
  \bibfield  {author} {\bibinfo {author} {\bibfnamefont {S.}~\bibnamefont
  {Hayden}}, \bibinfo {author} {\bibfnamefont {L.}~\bibnamefont {Taillefer}},
  \bibinfo {author} {\bibfnamefont {C.}~\bibnamefont {Vettier}}, \ and\
  \bibinfo {author} {\bibfnamefont {J.}~\bibnamefont {Flouquet}},\ }\bibfield
  {{Antiferromagnetic order in UPt$_3$ under pressure: Evidence for a direct
  coupling to Superconductivity}} {\emph {\bibinfo {title} {{Antiferromagnetic
  order in UPt$_3$ under pressure: Evidence for a direct coupling to
  Superconductivity}},\ }}\href {\doibase 10.1103/PhysRevB.46.8675} {\bibfield
  {journal} {\bibinfo  {journal} {Phys. Rev. B}\ }\textbf {\bibinfo {volume}
  {46}},\ \bibinfo {pages} {8675} (\bibinfo {year} {1992})}\BibitemShut
  {NoStop}%
\bibitem [{\citenamefont {Luk'yanchuk}(1991)}]{luk91}%
  \BibitemOpen
  \bibfield  {author} {\bibinfo {author} {\bibfnamefont {I.}~\bibnamefont
  {Luk'yanchuk}},\ }\bibfield  {{Superconducting kernel symmetry for an
  anisotropic superconductivity near H$_{c_2}$ and phase transitions in
  UPt$_3$}} {\emph {\bibinfo {title} {{Superconducting kernel symmetry for an
  anisotropic superconductivity near H$_{c_2}$ and phase transitions in
  UPt$_3$}},\ }}\href {\doibase 10.1051/jp1:1991197} {\bibfield  {journal}
  {\bibinfo  {journal} {J. de Phys. I}\ }\textbf {\bibinfo {volume} {1}},\
  \bibinfo {pages} {1155} (\bibinfo {year} {1991})}\BibitemShut {NoStop}%
\bibitem [{\citenamefont {Machida}\ and\ \citenamefont {Ozaki}(1991)}]{mac91}%
  \BibitemOpen
  \bibfield  {author} {\bibinfo {author} {\bibfnamefont {K.}~\bibnamefont
  {Machida}}\ and\ \bibinfo {author} {\bibfnamefont {M.}~\bibnamefont
  {Ozaki}},\ }\bibfield  {{Superconducting Double Transition in a Heavy Fermion
  Material UPt$_3$}} {\emph {\bibinfo {title} {{Superconducting Double
  Transition in a Heavy Fermion Material UPt$_3$}},\ }}\href {\doibase
  10.1103/PhysRevLett.66.3293} {\bibfield  {journal} {\bibinfo  {journal}
  {Phys. Rev. Lett.}\ }\textbf {\bibinfo {volume} {66}},\ \bibinfo {pages}
  {3293} (\bibinfo {year} {1991})}\BibitemShut {NoStop}%
\bibitem [{\citenamefont {Chen}\ and\ \citenamefont {Garg}(1993)}]{che93}%
  \BibitemOpen
  \bibfield  {author} {\bibinfo {author} {\bibfnamefont {D.}~\bibnamefont
  {Chen}}\ and\ \bibinfo {author} {\bibfnamefont {A.}~\bibnamefont {Garg}},\
  }\bibfield  {{Accidental Near Degeneracy of the Order Parameter for
  Superconducting UPt$_3$}} {\emph {\bibinfo {title} {{Accidental Near
  Degeneracy of the Order Parameter for Superconducting UPt$_3$}},\ }}\href
  {\doibase 10.1103/PhysRevLett.70.1689} {\bibfield  {journal} {\bibinfo
  {journal} {Phys. Rev. Lett.}\ }\textbf {\bibinfo {volume} {70}},\ \bibinfo
  {pages} {1689} (\bibinfo {year} {1993})}\BibitemShut {NoStop}%
\bibitem [{\citenamefont {Choi}\ and\ \citenamefont {Sauls}(1991)}]{cho91}%
  \BibitemOpen
  \bibfield  {author} {\bibinfo {author} {\bibfnamefont {C.}~\bibnamefont
  {Choi}}\ and\ \bibinfo {author} {\bibfnamefont {J.}~\bibnamefont {Sauls}},\
  }\bibfield  {{Identification of Odd-Parity Superconductivity in UPt$_3$ Based
  on Paramagnetic Limiting of the Upper Critical Field}} {\emph {\bibinfo
  {title} {{Identification of Odd-Parity Superconductivity in UPt$_3$ Based on
  Paramagnetic Limiting of the Upper Critical Field}},\ }}\href {\doibase
  10.1103/PhysRevLett.66.484} {\bibfield  {journal} {\bibinfo  {journal} {Phys.
  Rev. Lett.}\ }\textbf {\bibinfo {volume} {66}},\ \bibinfo {pages} {484}
  (\bibinfo {year} {1991})}\BibitemShut {NoStop}%
\bibitem [{\citenamefont {Sauls}(1994{\natexlab{c}})}]{sau94a}%
  \BibitemOpen
  \bibfield  {author} {\bibinfo {author} {\bibfnamefont {J.~A.}\ \bibnamefont
  {Sauls}},\ }\bibinfo {title} {{Fermi-Liquid Theory of Unconventional
  Superconductors}},\ in\ \href@noop {} {\emph {\bibinfo {booktitle} {{Strongly
  Correlated Electronic Materials: The Los Alamos Symposium 1993}}}},\ \bibinfo
  {editor} {edited by\ \bibinfo {editor} {\bibfnamefont {K.~S.}\ \bibnamefont
  {Bedell}}, \bibinfo {editor} {\bibfnamefont {Z.}~\bibnamefont {Wang}},
  \bibinfo {editor} {\bibfnamefont {D.~E.}\ \bibnamefont {Meltzer}}, \bibinfo
  {editor} {\bibfnamefont {A.~V.}\ \bibnamefont {Balatsky}}, \ and\ \bibinfo
  {editor} {\bibfnamefont {E.}~\bibnamefont {Abrahams}}}\ (\bibinfo
  {publisher} {Addison-Wesely},\ \bibinfo {address} {Reading, Mass.},\ \bibinfo
  {year} {1994})\ pp.\ \bibinfo {pages} {106--132},\ \Eprint
  {http://arxiv.org/abs/arXiv:2406.05230} {arXiv:2406.05230} \BibitemShut
  {NoStop}%
\bibitem [{\citenamefont {Shivaram}\ \emph {et~al.}(1986)\citenamefont
  {Shivaram}, \citenamefont {Jeong}, \citenamefont {Rosenbaum},\ and\
  \citenamefont {Hinks}}]{shi86a}%
  \BibitemOpen
  \bibfield  {author} {\bibinfo {author} {\bibfnamefont {B.}~\bibnamefont
  {Shivaram}}, \bibinfo {author} {\bibfnamefont {Y.}~\bibnamefont {Jeong}},
  \bibinfo {author} {\bibfnamefont {T.}~\bibnamefont {Rosenbaum}}, \ and\
  \bibinfo {author} {\bibfnamefont {D.}~\bibnamefont {Hinks}},\ }\bibfield
  {{Anisotropy of Transverse Sound in the Heavy Fermion Superconductor
  UPt$_3$}} {\emph {\bibinfo {title} {{Anisotropy of Transverse Sound in the
  Heavy Fermion Superconductor UPt$_3$}},\ }}\href {\doibase
  10.1103/PhysRevLett.56.1078} {\bibfield  {journal} {\bibinfo  {journal}
  {Phys. Rev. Lett.}\ }\textbf {\bibinfo {volume} {56}},\ \bibinfo {pages}
  {1078} (\bibinfo {year} {1986})}\BibitemShut {NoStop}%
\bibitem [{Note2()}]{Note2}%
  \BibitemOpen
  \bibinfo {note} {We use the ``energy representation'' in which the
  traditional variables $\protect \vec k$, $\protect \vec R$, $t$ of the
  distribution function are replaced by the equivalent set $\protect \vec k_f$,
  $\epsilon $, $\protect \vec R$, $t$. This amounts to a transformation to new
  coordinates in momentum space. The 3-dimensional momentum variable ($\protect
  \vec k$) is replaced by the 2-dimensional momentum variable $\protect \vec
  k_f$, which is the point on the Fermi surface nearest to $\protect \vec k$,
  and the 1-dimensional energy variable $\epsilon =E(\protect \vec k,\protect
  \vec R, t)-E_f$. The energy representation of the Boltzmann-Landau equation
  has a wide range of validity. It also describes overdamped excitations whose
  lifetime is comparable to or shorter than its oscillation period $\hbar
  /\epsilon $. For a review on transport theory in energy representation see
  Ref. (\protect \rev@citealpnum {ram86}).}\BibitemShut {Stop}%
\bibitem [{\citenamefont {Eliashberg}(1962)}]{eli62}%
  \BibitemOpen
  \bibfield  {author} {\bibinfo {author} {\bibfnamefont {G.~M.}\ \bibnamefont
  {Eliashberg}},\ }\bibfield  {{Transport Equation for a Degenerate System of
  Fermi Particles}} {\emph {\bibinfo {title} {{Transport Equation for a
  Degenerate System of Fermi Particles}},\ }}\href@noop {} {\bibfield
  {journal} {\bibinfo  {journal} {Sov. Phys. JETP}\ }\textbf {\bibinfo {volume}
  {14}},\ \bibinfo {pages} {886} (\bibinfo {year} {1962})},\ \bibinfo {note}
  {[ZhETF, 41, 1241, (1962)]}\BibitemShut {NoStop}%
\bibitem [{\citenamefont {Prange}\ and\ \citenamefont
  {Kadanoff}(1964)}]{pra64}%
  \BibitemOpen
  \bibfield  {author} {\bibinfo {author} {\bibfnamefont {R.~E.}\ \bibnamefont
  {Prange}}\ and\ \bibinfo {author} {\bibfnamefont {L.~P.}\ \bibnamefont
  {Kadanoff}},\ }\bibfield  {{Transport Theory for Electron-Phonon Interactions
  in Metals}} {\emph {\bibinfo {title} {{Transport Theory for Electron-Phonon
  Interactions in Metals}},\ }}\href {\doibase 10.1103/PhysRev.134.A566}
  {\bibfield  {journal} {\bibinfo  {journal} {Phys. Rev.}\ }\textbf {\bibinfo
  {volume} {134}},\ \bibinfo {pages} {A566} (\bibinfo {year}
  {1964})}\BibitemShut {NoStop}%
\bibitem [{Note3()}]{Note3}%
  \BibitemOpen
  \bibinfo {note} {The term {\protect \it large} means that the corresponding
  energies ($k_f^2/2m^{\ast } \sim m^{\ast }v_f^2/2 \sim E_f$) are much greater
  than typical excitation energies $\epsilon \sim k_BT_c$.}\BibitemShut {Stop}%
\bibitem [{\citenamefont {Baym}\ and\ \citenamefont {Pethick}(1991)}]{baym91}%
  \BibitemOpen
  \bibfield  {author} {\bibinfo {author} {\bibfnamefont {G.}~\bibnamefont
  {Baym}}\ and\ \bibinfo {author} {\bibfnamefont {C.~J.}\ \bibnamefont
  {Pethick}},\ }\href@noop {} {\emph {\bibinfo {title} {{Landau Fermi-Liquid
  Theory}}}}\ (\bibinfo  {publisher} {Wiley},\ \bibinfo {address} {New York},\
  \bibinfo {year} {1991})\BibitemShut {NoStop}%
\bibitem [{\citenamefont {Eilenberger}(1968)}]{eil68}%
  \BibitemOpen
  \bibfield  {author} {\bibinfo {author} {\bibfnamefont {G.}~\bibnamefont
  {Eilenberger}},\ }\bibfield  {{Transformation of Gorkov's Equation for Type
  II Superconductors into Transport-Like Equations}} {\emph {\bibinfo {title}
  {{Transformation of Gorkov's Equation for Type II Superconductors into
  Transport-Like Equations}},\ }}\href {\doibase 10.1007/BF01379803} {\bibfield
   {journal} {\bibinfo  {journal} {Zeit. f. Physik}\ }\textbf {\bibinfo
  {volume} {214}},\ \bibinfo {pages} {195} (\bibinfo {year}
  {1968})}\BibitemShut {NoStop}%
\bibitem [{\citenamefont {Larkin}\ and\ \citenamefont
  {Ovchinnikov}(1969)}]{lar69}%
  \BibitemOpen
  \bibfield  {author} {\bibinfo {author} {\bibfnamefont {A.~I.}\ \bibnamefont
  {Larkin}}\ and\ \bibinfo {author} {\bibfnamefont {Y.~N.}\ \bibnamefont
  {Ovchinnikov}},\ }\bibfield  {{Quasiclassical Method in the Theory of
  Superconductivity}} {\emph {\bibinfo {title} {{Quasiclassical Method in the
  Theory of Superconductivity}},\ }}\href@noop {} {\bibfield  {journal}
  {\bibinfo  {journal} {Sov. Phys. JETP}\ }\textbf {\bibinfo {volume} {28}},\
  \bibinfo {pages} {1200} (\bibinfo {year} {1969})}\BibitemShut {NoStop}%
\bibitem [{\citenamefont {Larkin}\ and\ \citenamefont
  {Ovchinnikov}(1975)}]{lar75}%
  \BibitemOpen
  \bibfield  {author} {\bibinfo {author} {\bibfnamefont {A.}~\bibnamefont
  {Larkin}}\ and\ \bibinfo {author} {\bibfnamefont {Y.}~\bibnamefont
  {Ovchinnikov}},\ }\bibfield  {{Nonlinear conductivity of superconductors in
  the mixed state}} {\emph {\bibinfo {title} {{Nonlinear conductivity of
  superconductors in the mixed state}},\ }}\href@noop {} {\bibfield  {journal}
  {\bibinfo  {journal} {Sov. Phys. JETP}\ }\textbf {\bibinfo {volume} {41}},\
  \bibinfo {pages} {960} (\bibinfo {year} {1975})}\BibitemShut {NoStop}%
\bibitem [{\citenamefont {Eliashberg}(1972)}]{eli72}%
  \BibitemOpen
  \bibfield  {author} {\bibinfo {author} {\bibfnamefont {G.~M.}\ \bibnamefont
  {Eliashberg}},\ }\bibfield  {{Inelastic electron collisions and
  nonequilibrium stationary states in superconductors}} {\emph {\bibinfo
  {title} {{Inelastic electron collisions and nonequilibrium stationary states
  in superconductors}},\ }}\href@noop {} {\bibfield  {journal} {\bibinfo
  {journal} {Sov. Phys. JETP}\ }\textbf {\bibinfo {volume} {34}},\ \bibinfo
  {pages} {668} (\bibinfo {year} {1972})}\BibitemShut {NoStop}%
\bibitem [{\citenamefont {Eckern}(1981)}]{eck81}%
  \BibitemOpen
  \bibfield  {author} {\bibinfo {author} {\bibfnamefont {U.}~\bibnamefont
  {Eckern}},\ }\bibfield  {{Quasiclassical approach to kinetic equations for
  superfluid $^3$He: General theory and application to the spin dynamics}}
  {\emph {\bibinfo {title} {{Quasiclassical approach to kinetic equations for
  superfluid $^3$He: General theory and application to the spin dynamics}},\
  }}\href {\doibase 10.1016/0003-4916(81)90256-6} {\bibfield  {journal}
  {\bibinfo  {journal} {Ann. Phys.}\ }\textbf {\bibinfo {volume} {133}},\
  \bibinfo {pages} {390} (\bibinfo {year} {1981})}\BibitemShut {NoStop}%
\bibitem [{\citenamefont {Serene}\ and\ \citenamefont {Rainer}(1983)}]{ser83}%
  \BibitemOpen
  \bibfield  {author} {\bibinfo {author} {\bibfnamefont {J.~W.}\ \bibnamefont
  {Serene}}\ and\ \bibinfo {author} {\bibfnamefont {D.}~\bibnamefont
  {Rainer}},\ }\bibfield  {{The Quasiclassical Approach to $^3He$}} {\emph
  {\bibinfo {title} {{The Quasiclassical Approach to $^3He$}},\ }}\href
  {\doibase 10.1016/0370-1573(83)90051-0} {\bibfield  {journal} {\bibinfo
  {journal} {Phys. Rep.}\ }\textbf {\bibinfo {volume} {101}},\ \bibinfo {pages}
  {221} (\bibinfo {year} {1983})}\BibitemShut {NoStop}%
\bibitem [{\citenamefont {Larkin}\ and\ \citenamefont
  {Ovchinnikov}(1986)}]{lar86}%
  \BibitemOpen
  \bibfield  {author} {\bibinfo {author} {\bibfnamefont {A.~I.}\ \bibnamefont
  {Larkin}}\ and\ \bibinfo {author} {\bibfnamefont {Y.~N.}\ \bibnamefont
  {Ovchinnikov}},\ }in\ \href@noop {} {\emph {\bibinfo {booktitle} {Modern
  Problems in Condensed Matter Physics}}},\ \bibinfo {editor} {edited by\
  \bibinfo {editor} {\bibfnamefont {D.}~\bibnamefont {Langenberg}}\ and\
  \bibinfo {editor} {\bibfnamefont {A.}~\bibnamefont {Larkin}}}\ (\bibinfo
  {publisher} {Elsevier Science Publishers},\ \bibinfo {address} {Amsterdam},\
  \bibinfo {year} {1986})\BibitemShut {NoStop}%
\bibitem [{\citenamefont {Rammer}\ and\ \citenamefont {Smith}(1986)}]{ram86}%
  \BibitemOpen
  \bibfield  {author} {\bibinfo {author} {\bibfnamefont {J.}~\bibnamefont
  {Rammer}}\ and\ \bibinfo {author} {\bibfnamefont {H.}~\bibnamefont {Smith}},\
  }\bibfield  {{Quantum field-theoretical methods in transport theory of
  metals}} {\emph {\bibinfo {title} {{Quantum field-theoretical methods in
  transport theory of metals}},\ }}\href {\doibase 10.1103/RevModPhys.58.323}
  {\bibfield  {journal} {\bibinfo  {journal} {Rev. Mod. Phys.}\ }\textbf
  {\bibinfo {volume} {58}},\ \bibinfo {pages} {323} (\bibinfo {year}
  {1986})}\BibitemShut {NoStop}%
\bibitem [{Note4()}]{Note4}%
  \BibitemOpen
  \bibinfo {note} {$n_{p(h)}(\protect \vec k_f,\protect \vec R;\epsilon
  ,t)d^2k_fd^3R\protect \, d\epsilon $ is the number of particle excitations
  (hole excitations) with momentum $\protect \vec k_f$, position $\protect \vec
  R$, and energy $\epsilon $ in the phase space element $d^2k_fd^3Rd\epsilon
  $.}\BibitemShut {Stop}%
\bibitem [{Note5()}]{Note5}%
  \BibitemOpen
  \bibinfo {note} {The amplitudes $\protect \mbox {\protect \underline
  {f}}^{K}$ and $\protect \vec {\protect \mbox {\protect \underline
  {f}}}^{\protect \,K}$ are redundant since they are related to $f^K$ and
  $\protect \vec {f}^K$ by fundamental symmetry relations.\cite
  {ser83}}\BibitemShut {NoStop}%
\bibitem [{Note6()}]{Note6}%
  \BibitemOpen
  \bibinfo {note} {The measured penetration depth is affected (dressed) by
  quasiparticle interactions.\cite {leg65} We follow the terminology of
  superfluid $^3$He and use the adjective ``stripped'' for properties of
  non-interacting quasiparticles.}\BibitemShut {Stop}%
\bibitem [{\citenamefont {Choi}\ and\ \citenamefont {Muzikar}(1988)}]{cho88}%
  \BibitemOpen
  \bibfield  {author} {\bibinfo {author} {\bibfnamefont {C.~H.}\ \bibnamefont
  {Choi}}\ and\ \bibinfo {author} {\bibfnamefont {P.}~\bibnamefont {Muzikar}},\
  }\bibfield  {{Superfluid Denity Tensor in Unconventional Superconductors}}
  {\emph {\bibinfo {title} {{Superfluid Denity Tensor in Unconventional
  Superconductors}},\ }}\href {\doibase 10.1103/PhysRevB.37.5947} {\bibfield
  {journal} {\bibinfo  {journal} {Phys. Rev. B}\ }\textbf {\bibinfo {volume}
  {37}},\ \bibinfo {pages} {5947} (\bibinfo {year} {1988})}\BibitemShut
  {NoStop}%
\bibitem [{\citenamefont {Choi}\ and\ \citenamefont
  {Muzikar}(1989{\natexlab{a}})}]{cho89b}%
  \BibitemOpen
  \bibfield  {author} {\bibinfo {author} {\bibfnamefont {C.}~\bibnamefont
  {Choi}}\ and\ \bibinfo {author} {\bibfnamefont {P.}~\bibnamefont {Muzikar}},\
  }\bibfield  {Theory of the superfluid density tensor in unconventional
  superconductors: Impurity scattering and band-structure effects} {\emph
  {\bibinfo {title} {Theory of the superfluid density tensor in unconventional
  superconductors: Impurity scattering and band-structure effects},\ }}\href
  {\doibase 10.1103/PhysRevB.39.11296} {\bibfield  {journal} {\bibinfo
  {journal} {Phys. Rev. B}\ }\textbf {\bibinfo {volume} {39}},\ \bibinfo
  {pages} {11296} (\bibinfo {year} {1989}{\natexlab{a}})}\BibitemShut {NoStop}%
\bibitem [{\citenamefont {Tokuyasu}\ \emph {et~al.}(1990)\citenamefont
  {Tokuyasu}, \citenamefont {Hess},\ and\ \citenamefont {Sauls}}]{tok90}%
  \BibitemOpen
  \bibfield  {author} {\bibinfo {author} {\bibfnamefont {T.~A.}\ \bibnamefont
  {Tokuyasu}}, \bibinfo {author} {\bibfnamefont {D.~W.}\ \bibnamefont {Hess}},
  \ and\ \bibinfo {author} {\bibfnamefont {J.~A.}\ \bibnamefont {Sauls}},\
  }\bibfield  {{Vortex states in an unconventional superconductor and the mixed
  phases of $UPt3$}} {\emph {\bibinfo {title} {{Vortex states in an
  unconventional superconductor and the mixed phases of $UPt3$}},\ }}\href
  {\doibase 10.1103/PhysRevB.41.8891} {\bibfield  {journal} {\bibinfo
  {journal} {Phys. Rev. B}\ }\textbf {\bibinfo {volume} {41}},\ \bibinfo
  {pages} {8891} (\bibinfo {year} {1990})}\BibitemShut {NoStop}%
\bibitem [{\citenamefont {Tokuyasu}\ and\ \citenamefont
  {Sauls}(1990)}]{tok90a}%
  \BibitemOpen
  \bibfield  {author} {\bibinfo {author} {\bibfnamefont {T.}~\bibnamefont
  {Tokuyasu}}\ and\ \bibinfo {author} {\bibfnamefont {J.}~\bibnamefont
  {Sauls}},\ }\bibfield  {Stability of doubly quantized vortices in
  unconventional superconductors} {\emph {\bibinfo {title} {Stability of doubly
  quantized vortices in unconventional superconductors},\ }}\href {\doibase
  10.1016/S0921-4526(90)81023-H} {\bibfield  {journal} {\bibinfo  {journal}
  {Physica B: Cond. Matt.}\ }\textbf {\bibinfo {volume} {165-166}},\ \bibinfo
  {pages} {347} (\bibinfo {year} {1990})},\ \bibinfo {note} {proceedings of the
  19th International Conference on Low Temperature Physics}\BibitemShut
  {NoStop}%
\bibitem [{\citenamefont {Palumbo}\ \emph
  {et~al.}(1990{\natexlab{a}})\citenamefont {Palumbo}, \citenamefont
  {Muzikar},\ and\ \citenamefont {Sauls}}]{pal90}%
  \BibitemOpen
  \bibfield  {author} {\bibinfo {author} {\bibfnamefont {M.}~\bibnamefont
  {Palumbo}}, \bibinfo {author} {\bibfnamefont {P.}~\bibnamefont {Muzikar}}, \
  and\ \bibinfo {author} {\bibfnamefont {J.}~\bibnamefont {Sauls}},\ }\bibfield
   {{Magnetic Instabilities in Unconventional Superconductors}} {\emph
  {\bibinfo {title} {{Magnetic Instabilities in Unconventional
  Superconductors}},\ }}\href {\doibase 10.1103/PhysRevB.42.2681} {\bibfield
  {journal} {\bibinfo  {journal} {Phys. Rev. B}\ }\textbf {\bibinfo {volume}
  {42}},\ \bibinfo {pages} {2681} (\bibinfo {year}
  {1990}{\natexlab{a}})}\BibitemShut {NoStop}%
\bibitem [{\citenamefont {Barash}\ and\ \citenamefont
  {Mel'nikov}(1991)}]{bar91}%
  \BibitemOpen
  \bibfield  {author} {\bibinfo {author} {\bibfnamefont {Y.}~\bibnamefont
  {Barash}}\ and\ \bibinfo {author} {\bibfnamefont {A.~S.}\ \bibnamefont
  {Mel'nikov}},\ }\bibfield  {{Internal Structure of Vortices in Exotic
  Superconductors Near the Lower Critical Field}} {\emph {\bibinfo {title}
  {{Internal Structure of Vortices in Exotic Superconductors Near the Lower
  Critical Field}},\ }}\href@noop {} {\bibfield  {journal} {\bibinfo  {journal}
  {Sov. Phys. JETP}\ }\textbf {\bibinfo {volume} {73}},\ \bibinfo {pages} {170}
  (\bibinfo {year} {1991})}\BibitemShut {NoStop}%
\bibitem [{\citenamefont {Palumbo}\ and\ \citenamefont
  {Muzikar}(1992)}]{pal92}%
  \BibitemOpen
  \bibfield  {author} {\bibinfo {author} {\bibfnamefont {M.}~\bibnamefont
  {Palumbo}}\ and\ \bibinfo {author} {\bibfnamefont {P.}~\bibnamefont
  {Muzikar}},\ }\bibfield  {{New ground states in unconventional
  superconductors: Broken translational and time-reversal symmetry}} {\emph
  {\bibinfo {title} {{New ground states in unconventional superconductors:
  Broken translational and time-reversal symmetry}},\ }}\href {\doibase
  10.1103/PhysRevB.45.12620} {\bibfield  {journal} {\bibinfo  {journal} {Phys.
  Rev. B}\ }\textbf {\bibinfo {volume} {45}},\ \bibinfo {pages} {12620}
  (\bibinfo {year} {1992})}\BibitemShut {NoStop}%
\bibitem [{\citenamefont {Mel'nikov}(1992)}]{mel92}%
  \BibitemOpen
  \bibfield  {author} {\bibinfo {author} {\bibfnamefont {A.~S.}\ \bibnamefont
  {Mel'nikov}},\ }\bibfield  {{Phase Transitions in Vortex Lattices of
  Hexagonal Exotic Superconductors}} {\emph {\bibinfo {title} {{Phase
  Transitions in Vortex Lattices of Hexagonal Exotic Superconductors}},\
  }}\href@noop {} {\bibfield  {journal} {\bibinfo  {journal} {Sov. Phys. JETP}\
  }\textbf {\bibinfo {volume} {74}},\ \bibinfo {pages} {1059} (\bibinfo {year}
  {1992})}\BibitemShut {NoStop}%
\bibitem [{\citenamefont {{P. Hirschfeld and P. W\"olfle and J. A. Sauls and D.
  Einzel and W.O. Putikka}}(1989)}]{hir89}%
  \BibitemOpen
  \bibfield  {author} {\bibinfo {author} {\bibnamefont {{P. Hirschfeld and P.
  W\"olfle and J. A. Sauls and D. Einzel and W.O. Putikka}}},\ }\bibfield
  {{Electromagnetic Absorption in Anisotropic Superconductors}} {\emph
  {\bibinfo {title} {{Electromagnetic Absorption in Anisotropic
  Superconductors}},\ }}\href {\doibase 10.1103/PhysRevB.40.6695} {\bibfield
  {journal} {\bibinfo  {journal} {Phys. Rev. B}\ }\textbf {\bibinfo {volume}
  {40}},\ \bibinfo {pages} {6695} (\bibinfo {year} {1989})}\BibitemShut
  {NoStop}%
\bibitem [{\citenamefont {Yip}\ and\ \citenamefont {Sauls}(1992)}]{yip92}%
  \BibitemOpen
  \bibfield  {author} {\bibinfo {author} {\bibfnamefont {S.~K.}\ \bibnamefont
  {Yip}}\ and\ \bibinfo {author} {\bibfnamefont {J.~A.}\ \bibnamefont
  {Sauls}},\ }\bibfield  {{Circular Dichroism and Birefringence in
  Unconventional Superconductors}} {\emph {\bibinfo {title} {{Circular
  Dichroism and Birefringence in Unconventional Superconductors}},\ }}\href
  {\doibase 10.1007/BF01151804} {\bibfield  {journal} {\bibinfo  {journal} {J.
  Low Temp. Phys.}\ }\textbf {\bibinfo {volume} {86}},\ \bibinfo {pages} {257}
  (\bibinfo {year} {1992})}\BibitemShut {NoStop}%
\bibitem [{\citenamefont {Ambegaokar}\ \emph {et~al.}(1974)\citenamefont
  {Ambegaokar}, \citenamefont {{de Gennes}},\ and\ \citenamefont
  {Rainer}}]{amb74}%
  \BibitemOpen
  \bibfield  {author} {\bibinfo {author} {\bibfnamefont {V.}~\bibnamefont
  {Ambegaokar}}, \bibinfo {author} {\bibfnamefont {P.}~\bibnamefont {{de
  Gennes}}}, \ and\ \bibinfo {author} {\bibfnamefont {D.}~\bibnamefont
  {Rainer}},\ }\bibfield  {{Landau-Ginzburg Equations for an Anisotropic
  Superfluid}} {\emph {\bibinfo {title} {{Landau-Ginzburg Equations for an
  Anisotropic Superfluid}},\ }}\href {\doibase 10.1103/PhysRevA.9.2676}
  {\bibfield  {journal} {\bibinfo  {journal} {Phys. Rev. A}\ }\textbf {\bibinfo
  {volume} {9}},\ \bibinfo {pages} {2676} (\bibinfo {year} {1974})}\BibitemShut
  {NoStop}%
\bibitem [{\citenamefont {Kurkij\"arvi}\ \emph {et~al.}(1987)\citenamefont
  {Kurkij\"arvi}, \citenamefont {Rainer},\ and\ \citenamefont {Sauls}}]{kur87}%
  \BibitemOpen
  \bibfield  {author} {\bibinfo {author} {\bibfnamefont {J.}~\bibnamefont
  {Kurkij\"arvi}}, \bibinfo {author} {\bibfnamefont {D.}~\bibnamefont
  {Rainer}}, \ and\ \bibinfo {author} {\bibfnamefont {J.~A.}\ \bibnamefont
  {Sauls}},\ }\bibfield  {{Superfluid {$^3$He} and Heavy Fermion
  Superconductors Near Surfaces and Interfaces}} {\emph {\bibinfo {title}
  {{Superfluid {$^3$He} and Heavy Fermion Superconductors Near Surfaces and
  Interfaces}},\ }}\href {\doibase 10.1139/p87-227} {\bibfield  {journal}
  {\bibinfo  {journal} {Can. J. Phys.}\ }\textbf {\bibinfo {volume} {65}},\
  \bibinfo {pages} {1440} (\bibinfo {year} {1987})}\BibitemShut {NoStop}%
\bibitem [{\citenamefont {Geshkenbein}\ and\ \citenamefont
  {Larkin}(1986)}]{ges85}%
  \BibitemOpen
  \bibfield  {author} {\bibinfo {author} {\bibfnamefont {V.~B.}\ \bibnamefont
  {Geshkenbein}}\ and\ \bibinfo {author} {\bibfnamefont {A.~I.}\ \bibnamefont
  {Larkin}},\ }\bibfield  {{The Josephson effect in superconductors with heavy
  fermions}} {\emph {\bibinfo {title} {{The Josephson effect in superconductors
  with heavy fermions}},\ }}\href@noop {} {\bibfield  {journal} {\bibinfo
  {journal} {Sov. Phys. JETP Lett.}\ }\textbf {\bibinfo {volume} {43}},\
  \bibinfo {pages} {395} (\bibinfo {year} {1986})}\BibitemShut {NoStop}%
\bibitem [{\citenamefont {Millis}\ \emph {et~al.}(1988)\citenamefont {Millis},
  \citenamefont {Rainer},\ and\ \citenamefont {Sauls}}]{mil88}%
  \BibitemOpen
  \bibfield  {author} {\bibinfo {author} {\bibfnamefont {A.}~\bibnamefont
  {Millis}}, \bibinfo {author} {\bibfnamefont {D.}~\bibnamefont {Rainer}}, \
  and\ \bibinfo {author} {\bibfnamefont {J.~A.}\ \bibnamefont {Sauls}},\
  }\bibfield  {{Quasiclassical Theory of Superconductivity Near Magnetically
  Active Interfaces}} {\emph {\bibinfo {title} {{Quasiclassical Theory of
  Superconductivity Near Magnetically Active Interfaces}},\ }}\href {\doibase
  10.1103/PhysRevB.38.4504} {\bibfield  {journal} {\bibinfo  {journal} {Phys.
  Rev. B}\ }\textbf {\bibinfo {volume} {38}},\ \bibinfo {pages} {4504}
  (\bibinfo {year} {1988})}\BibitemShut {NoStop}%
\bibitem [{\citenamefont {Yip}(1993)}]{yip93}%
  \BibitemOpen
  \bibfield  {author} {\bibinfo {author} {\bibfnamefont {S.}~\bibnamefont
  {Yip}},\ }\bibfield  {{Weak Link between Conventional and Unconventional
  Superconductors}} {\emph {\bibinfo {title} {{Weak Link between Conventional
  and Unconventional Superconductors}},\ }}\href {\doibase 10.1007/BF00120849}
  {\bibfield  {journal} {\bibinfo  {journal} {J. Low Temp. Phys.}\ }\textbf
  {\bibinfo {volume} {91}},\ \bibinfo {pages} {203} (\bibinfo {year}
  {1993})}\BibitemShut {NoStop}%
\bibitem [{\citenamefont {Yip}\ \emph {et~al.}(1994)\citenamefont {Yip},
  \citenamefont {Sun},\ and\ \citenamefont {Sauls}}]{yip94}%
  \BibitemOpen
  \bibfield  {author} {\bibinfo {author} {\bibfnamefont {S.}~\bibnamefont
  {Yip}}, \bibinfo {author} {\bibfnamefont {Y.~S.}\ \bibnamefont {Sun}}, \ and\
  \bibinfo {author} {\bibfnamefont {J.~A.}\ \bibnamefont {Sauls}},\ }\bibfield
  {{Josephson Effects in Superconducting Conventional/Unconventional Tunnel
  Junctions and Weak Links}} {\emph {\bibinfo {title} {{Josephson Effects in
  Superconducting Conventional/Unconventional Tunnel Junctions and Weak
  Links}},\ }}\href {\doibase 10.1016/0921-4526(94)91484-2} {\bibfield
  {journal} {\bibinfo  {journal} {Physica B}\ }\textbf {\bibinfo {volume}
  {194-196}},\ \bibinfo {pages} {1969} (\bibinfo {year} {1994})}\BibitemShut
  {NoStop}%
\bibitem [{\citenamefont {Wollman}\ \emph {et~al.}(1994)\citenamefont
  {Wollman}, \citenamefont {Harlingen}, \citenamefont {Lee}, \citenamefont
  {Ginsberg},\ and\ \citenamefont {Leggett}}]{van94}%
  \BibitemOpen
  \bibfield  {author} {\bibinfo {author} {\bibfnamefont {D.}~\bibnamefont
  {Wollman}}, \bibinfo {author} {\bibfnamefont {D.~V.}\ \bibnamefont
  {Harlingen}}, \bibinfo {author} {\bibfnamefont {W.}~\bibnamefont {Lee}},
  \bibinfo {author} {\bibfnamefont {D.}~\bibnamefont {Ginsberg}}, \ and\
  \bibinfo {author} {\bibfnamefont {A.}~\bibnamefont {Leggett}},\ }\bibfield
  {{Phase coherence in YBCO-Pb SQUIDs: Direct experimental determination of the
  symmetry of the pairing state in high temperature superconductors}} {\emph
  {\bibinfo {title} {{Phase coherence in YBCO-Pb SQUIDs: Direct experimental
  determination of the symmetry of the pairing state in high temperature
  superconductors}},\ }}\href {\doibase 10.1016/0921-4526(94)91334-X}
  {\bibfield  {journal} {\bibinfo  {journal} {Physica B}\ }\textbf {\bibinfo
  {volume} {194-196}},\ \bibinfo {pages} {1669} (\bibinfo {year} {1994})},\
  \bibinfo {note} {proceedings of LT20}\BibitemShut {NoStop}%
\bibitem [{\citenamefont {Vollhardt}\ and\ \citenamefont
  {W\"olfle}(1990)}]{vollhardt90}%
  \BibitemOpen
  \bibfield  {author} {\bibinfo {author} {\bibfnamefont {D.}~\bibnamefont
  {Vollhardt}}\ and\ \bibinfo {author} {\bibfnamefont {P.}~\bibnamefont
  {W\"olfle}},\ }\href@noop {} {\emph {\bibinfo {title} {{The Superfluid Phases
  of $^3$He}}}}\ (\bibinfo  {publisher} {Taylor \& Francis},\ \bibinfo
  {address} {New York},\ \bibinfo {year} {1990})\BibitemShut {NoStop}%
\bibitem [{\citenamefont {Mc{K}enzie}\ and\ \citenamefont
  {Sauls}(1990)}]{mck90}%
  \BibitemOpen
  \bibfield  {author} {\bibinfo {author} {\bibfnamefont {R.~H.}\ \bibnamefont
  {Mc{K}enzie}}\ and\ \bibinfo {author} {\bibfnamefont {J.~A.}\ \bibnamefont
  {Sauls}},\ }\bibinfo {title} {{Collective Modes and Nonlinear Acoustics in
  Superfluid $^3He$-B}},\ in\ \href@noop {} {\emph {\bibinfo {booktitle}
  {Helium Three}}},\ \bibinfo {editor} {edited by\ \bibinfo {editor}
  {\bibfnamefont {W.~P.}\ \bibnamefont {Halperin}}\ and\ \bibinfo {editor}
  {\bibfnamefont {L.~P.}\ \bibnamefont {Pitaevskii}}}\ (\bibinfo  {publisher}
  {Elsevier Science Publishers},\ \bibinfo {address} {Amsterdam},\ \bibinfo
  {year} {1990})\ p.\ \bibinfo {pages} {255},\ \Eprint
  {http://arxiv.org/abs/1309.6018} {1309.6018} \BibitemShut {NoStop}%
\bibitem [{\citenamefont {Ueda}\ and\ \citenamefont {Rice}(1985)}]{ued85}%
  \BibitemOpen
  \bibfield  {author} {\bibinfo {author} {\bibfnamefont {K.}~\bibnamefont
  {Ueda}}\ and\ \bibinfo {author} {\bibfnamefont {T.}~\bibnamefont {Rice}},\
  }\bibfield  {p-wave superconductivity in cubic metals} {\emph {\bibinfo
  {title} {p-wave superconductivity in cubic metals},\ }}\href {\doibase
  10.1103/PhysRevB.31.7114} {\bibfield  {journal} {\bibinfo  {journal} {Phys.
  Rev. B}\ }\textbf {\bibinfo {volume} {31}},\ \bibinfo {pages} {7114}
  (\bibinfo {year} {1985})}\BibitemShut {NoStop}%
\bibitem [{\citenamefont {Fetter}(1965)}]{fet65}%
  \BibitemOpen
  \bibfield  {author} {\bibinfo {author} {\bibfnamefont {A.~L.}\ \bibnamefont
  {Fetter}},\ }\bibfield  {{Spherical Impurity in an Infinite Superconductor}}
  {\emph {\bibinfo {title} {{Spherical Impurity in an Infinite
  Superconductor}},\ }}\href {\doibase 10.1103/PhysRev.140.A1921} {\bibfield
  {journal} {\bibinfo  {journal} {Phys. Rev.}\ }\textbf {\bibinfo {volume}
  {140}},\ \bibinfo {pages} {1921} (\bibinfo {year} {1965})}\BibitemShut
  {NoStop}%
\bibitem [{\citenamefont {Rainer}\ and\ \citenamefont {Vuorio}(1977)}]{rai77}%
  \BibitemOpen
  \bibfield  {author} {\bibinfo {author} {\bibfnamefont {D.}~\bibnamefont
  {Rainer}}\ and\ \bibinfo {author} {\bibfnamefont {M.}~\bibnamefont
  {Vuorio}},\ }\bibfield  {{Small Objects in Superfluid $^3$He}} {\emph
  {\bibinfo {title} {{Small Objects in Superfluid $^3$He}},\ }}\href {\doibase
  10.1088/0022-3719/10/16/018} {\bibfield  {journal} {\bibinfo  {journal} {J.
  Phys. C}\ }\textbf {\bibinfo {volume} {10}},\ \bibinfo {pages} {3093}
  (\bibinfo {year} {1977})}\BibitemShut {NoStop}%
\bibitem [{\citenamefont {Choi}\ and\ \citenamefont
  {Muzikar}(1989{\natexlab{b}})}]{cho89a}%
  \BibitemOpen
  \bibfield  {author} {\bibinfo {author} {\bibfnamefont {C.}~\bibnamefont
  {Choi}}\ and\ \bibinfo {author} {\bibfnamefont {P.}~\bibnamefont {Muzikar}},\
  }\bibfield  {{Impurity-induced Magnetic Fields in Unconventional
  Superconductors}} {\emph {\bibinfo {title} {{Impurity-induced Magnetic Fields
  in Unconventional Superconductors}},\ }}\href {\doibase
  10.1103/PhysRevB.39.9664} {\bibfield  {journal} {\bibinfo  {journal} {Phys.
  Rev. B}\ }\textbf {\bibinfo {volume} {39}},\ \bibinfo {pages} {9664(R)}
  (\bibinfo {year} {1989}{\natexlab{b}})}\BibitemShut {NoStop}%
\bibitem [{\citenamefont {Choi}\ and\ \citenamefont
  {Muzikar}(1989{\natexlab{c}})}]{cho89}%
  \BibitemOpen
  \bibfield  {author} {\bibinfo {author} {\bibfnamefont {C.}~\bibnamefont
  {Choi}}\ and\ \bibinfo {author} {\bibfnamefont {P.}~\bibnamefont {Muzikar}},\
  }\bibfield  {{Gradient free energy and intrinsic angular momentum in
  unconventional superconductors}} {\emph {\bibinfo {title} {{Gradient free
  energy and intrinsic angular momentum in unconventional superconductors}},\
  }}\href {\doibase 10.1103/PhysRevB.40.5144} {\bibfield  {journal} {\bibinfo
  {journal} {Phys. Rev. B}\ }\textbf {\bibinfo {volume} {40}},\ \bibinfo
  {pages} {5144} (\bibinfo {year} {1989}{\natexlab{c}})}\BibitemShut {NoStop}%
\bibitem [{\citenamefont {Palumbo}\ \emph
  {et~al.}(1990{\natexlab{b}})\citenamefont {Palumbo}, \citenamefont {Choi},\
  and\ \citenamefont {Muzikar}}]{pal90a}%
  \BibitemOpen
  \bibfield  {author} {\bibinfo {author} {\bibfnamefont {M.}~\bibnamefont
  {Palumbo}}, \bibinfo {author} {\bibfnamefont {C.~H.}\ \bibnamefont {Choi}}, \
  and\ \bibinfo {author} {\bibfnamefont {P.}~\bibnamefont {Muzikar}},\
  }\bibfield  {{Ginzburg-Landau free energy in unconventional superconductors}}
  {\emph {\bibinfo {title} {{Ginzburg-Landau free energy in unconventional
  superconductors}},\ }}\href {\doibase 10.1016/S0921-4526(09)80133-6}
  {\bibfield  {journal} {\bibinfo  {journal} {Physica B}\ }\textbf {\bibinfo
  {volume} {165-166}},\ \bibinfo {pages} {1095} (\bibinfo {year}
  {1990}{\natexlab{b}})}\BibitemShut {NoStop}%
\bibitem [{\citenamefont {Cross}(1975)}]{cro75}%
  \BibitemOpen
  \bibfield  {author} {\bibinfo {author} {\bibfnamefont {M.~C.}\ \bibnamefont
  {Cross}},\ }\bibfield  {{A Generalized Ginzburg-Landau Approach to the
  Superfluidity of $^3$He}} {\emph {\bibinfo {title} {{A Generalized
  Ginzburg-Landau Approach to the Superfluidity of $^3$He}},\ }}\href {\doibase
  10.1007/BF01141607} {\bibfield  {journal} {\bibinfo  {journal} {J. Low Temp.
  Phys.}\ }\textbf {\bibinfo {volume} {21}},\ \bibinfo {pages} {525} (\bibinfo
  {year} {1975})}\BibitemShut {NoStop}%
\bibitem [{Note7()}]{Note7}%
  \BibitemOpen
  \bibinfo {note} {Similar calculations to those in this section can be carried
  out for the E$_{2u}$ representation, which in our opinion is the most likely
  candidate for the phases of UPt$_3$.}\BibitemShut {Stop}%
\bibitem [{\citenamefont {Mermin}\ and\ \citenamefont {Muzikar}(1980)}]{mer80}%
  \BibitemOpen
  \bibfield  {author} {\bibinfo {author} {\bibfnamefont {N.}~\bibnamefont
  {Mermin}}\ and\ \bibinfo {author} {\bibfnamefont {P.}~\bibnamefont
  {Muzikar}},\ }\bibfield  {{Cooper pairs versus Bose condensed molecules: The
  ground-state current in superfluid $^3$He}} {\emph {\bibinfo {title} {{Cooper
  pairs versus Bose condensed molecules: The ground-state current in superfluid
  $^3$He}},\ }}\href {\doibase 10.1103/PhysRevB.21.980} {\bibfield  {journal}
  {\bibinfo  {journal} {Phys. Rev. B}\ }\textbf {\bibinfo {volume} {21}},\
  \bibinfo {pages} {980} (\bibinfo {year} {1980})}\BibitemShut {NoStop}%
\bibitem [{\citenamefont {Muzikar}(1991)}]{muz91}%
  \BibitemOpen
  \bibfield  {author} {\bibinfo {author} {\bibfnamefont {P.}~\bibnamefont
  {Muzikar}},\ }\bibfield  {{Local structure of supercurrent flow past an
  impurity}} {\emph {\bibinfo {title} {{Local structure of supercurrent flow
  past an impurity}},\ }}\href {\doibase 10.1103/PhysRevB.43.10201} {\bibfield
  {journal} {\bibinfo  {journal} {Phys. Rev. B}\ }\textbf {\bibinfo {volume}
  {43}},\ \bibinfo {pages} {10201} (\bibinfo {year} {1991})}\BibitemShut
  {NoStop}%
\bibitem [{\citenamefont {Thuneberg}\ \emph {et~al.}(1981)\citenamefont
  {Thuneberg}, \citenamefont {Kurkij\"arvi},\ and\ \citenamefont
  {Rainer}}]{thu81}%
  \BibitemOpen
  \bibfield  {author} {\bibinfo {author} {\bibfnamefont {E.}~\bibnamefont
  {Thuneberg}}, \bibinfo {author} {\bibfnamefont {J.}~\bibnamefont
  {Kurkij\"arvi}}, \ and\ \bibinfo {author} {\bibfnamefont {D.}~\bibnamefont
  {Rainer}},\ }\bibfield  {{Quasiclassical Theory of Ions in $^3$He}} {\emph
  {\bibinfo {title} {{Quasiclassical Theory of Ions in $^3$He}},\ }}\href
  {\doibase 10.1088/0022-3719/14/36/006} {\bibfield  {journal} {\bibinfo
  {journal} {J. Phys. C}\ }\textbf {\bibinfo {volume} {14}},\ \bibinfo {pages}
  {5615} (\bibinfo {year} {1981})}\BibitemShut {NoStop}%
\bibitem [{\citenamefont {Thuneberg}\ \emph {et~al.}(1984)\citenamefont
  {Thuneberg}, \citenamefont {Kurkij\"arvi},\ and\ \citenamefont
  {Rainer}}]{thu84}%
  \BibitemOpen
  \bibfield  {author} {\bibinfo {author} {\bibfnamefont {E.}~\bibnamefont
  {Thuneberg}}, \bibinfo {author} {\bibfnamefont {J.}~\bibnamefont
  {Kurkij\"arvi}}, \ and\ \bibinfo {author} {\bibfnamefont {D.}~\bibnamefont
  {Rainer}},\ }\bibfield  {{Elementary-flux-pinning potential in type-II
  superconductors}} {\emph {\bibinfo {title} {{Elementary-flux-pinning
  potential in type-II superconductors}},\ }}\href {\doibase
  10.1103/PhysRevB.29.3913} {\bibfield  {journal} {\bibinfo  {journal} {Phys.
  Rev. B}\ }\textbf {\bibinfo {volume} {29}},\ \bibinfo {pages} {3913}
  (\bibinfo {year} {1984})}\BibitemShut {NoStop}%
\bibitem [{\citenamefont {Choi}\ and\ \citenamefont {Muzikar}(1990)}]{cho90}%
  \BibitemOpen
  \bibfield  {author} {\bibinfo {author} {\bibfnamefont {C.}~\bibnamefont
  {Choi}}\ and\ \bibinfo {author} {\bibfnamefont {P.}~\bibnamefont {Muzikar}},\
  }\bibfield  {{Impurity Magnetic Moments in Unconventional Superconductors}}
  {\emph {\bibinfo {title} {{Impurity Magnetic Moments in Unconventional
  Superconductors}},\ }}\href {\doibase 10.1103/PhysRevB.41.1812} {\bibfield
  {journal} {\bibinfo  {journal} {Phys. Rev. B}\ }\textbf {\bibinfo {volume}
  {41}},\ \bibinfo {pages} {1812} (\bibinfo {year} {1990})}\BibitemShut
  {NoStop}%
\bibitem [{\citenamefont {Leggett}(1965)}]{leg65}%
  \BibitemOpen
  \bibfield  {author} {\bibinfo {author} {\bibfnamefont {A.~J.}\ \bibnamefont
  {Leggett}},\ }\bibfield  {{Theory of a Superfluid Fermi Liquid I: General
  Formalism and Static Properties}} {\emph {\bibinfo {title} {{Theory of a
  Superfluid Fermi Liquid I: General Formalism and Static Properties}},\
  }}\href {\doibase 10.1103/PhysRev.140.A1869} {\bibfield  {journal} {\bibinfo
  {journal} {Phys. Rev.}\ }\textbf {\bibinfo {volume} {140}},\ \bibinfo {pages}
  {A1869} (\bibinfo {year} {1965})}\BibitemShut {NoStop}%
\end{thebibliography}

%
\end{document}